# Possibility of coherent electron transport in a nanoscale circuit

Mark J. Hagmann

NewPath Research L.L.C., 2880 S. Main Street, Suite 214, Salt Lake City, Utah USA 84115

September 25, 2020

## ABSTRACT

Others have solved the Schrödinger equation to estimate the tunneling current between two electrodes at specified potentials, or the transmission through a potential barrier, assuming that an incident wave causes one reflected wave and one transmitted wave. However, this may not be appropriate in some nanoscale circuits because the electron mean-free path may be as long as 68 nm in metals. Thus, the wavefunction may be coherent throughout the metal components in a circuit if the interaction of the electrons with the surface of conductors and grain boundaries, which reduces the mean-free path, is reduced. We consider the use of single-crystal wires, and include a tunneling junction to focus and collimate the electrons near the axis, to further reduce their interaction with the surface of the wire. Our simulations suggest that, in addition to the incoherent phenomena, there are extremely sharply-defined coherent modes in nanoscale circuits. Algorithms are presented with examples to determine the sets of the parameters for these modes. Other algorithms are presented to determine the normalized coefficients in the wavefunction and the distribution of current in the circuits. This is done using only algebra with calculus for analytical solutions of the Schrödinger equation.

## I. INTRODUCTION

In 1991 Kalotas and Lee [1] introduced the transfer-matrix method to solve the one-dimensional Schrödinger equation when modeling quantum tunneling in arbitrary static potential barriers. Grossel, Vigoureux and Baida [2] compared the stability of this method with that when using WKB. We were the first to apply the transfer-matrix method with a time-dependent potential and modeled the effects of the barrier traversal time in laser-assisted scanning tunneling microscopy [3]. These simulations guided our development of microwave oscillators that are based on laser-assisted field emission [4] and the generation of microwave frequency combs by focusing a mode-locked laser on the tunneling junction of a scanning tunneling microscope (STM) [5]. Now we are studying a variant of the STM where extremely low-power ($\approx$ 3 aW) microwave harmonics of the laser pulse repetition rate have a signal-to-noise ratio of 20 dB. These extremely low-noise measurements are possible because the quality factor (Q) at each harmonic is $10^{12}$ which is five times that of cryogenic microwave cavities [6]. Basing feedback control of the tip-sample separation in an STM on these harmonics instead of the high-noise tunneling current can increase the speed and stability of imaging and does not require the continuous intense static field ($\approx$ 1 V/nm) that causes electroporation, as well as damage, in biological samples and band-bending in semiconductors when using an STM [7]. The present analysis is part of our effort to develop a macroscopic instrument coupled to a nanoscale circuit for advanced scanning probe microscopy.

## II. DELIMITATIONS

We model a tunneling junction that has two ideal metal electrodes with the same work function and is coupled to a circuit with extremely low electrical resistivity but it is not a superconductor. The distribution of electron energies and the effects of images of the tunneling electrons at each electrode are neglected to obtain analytical solutions. We acknowledge that the Dirac equation would more accurately simulate the properties of the electron (Per-Olov Löwdin, personal communication, 1998) but at present we continue to use the Schrödinger equation.

## III. FOCUS AND COLUMNATE ELECTRONS WITH A TUNNELING JUNCTION.

Gall simulated the electron mean-free path $\lambda$ for 20 metallic elements that have different bulk resistivities $\rho_0$ at room temperature to show that $\lambda$ is greatest at 68.2 nm for beryllium [8]. For comparison, these calculations show that $\lambda$ is 53.3 nm for silver, 39.9 nm for copper, and 37.7 nm for gold. However, the apparent bulk electrical resistivity $\rho'$ of a metal wire increases as the diameter is reduced because of increased scattering of the electrons at the surfaces and grain boundaries [8],[9]. Thus, for fine wires the apparent resistivity $\rho'$ is greater than $\rho_0$ and proportional to the product $\rho_0\lambda$ [8]. For example, Rhenium, with a bulk resistivity that is 2.1 times that of copper, has an apparent resistivity that is 0.48 times that of copper.

We consider the case where electrons are injected axially into a small area at the end of a cylindrical wire. This may be done by using a tunneling junction because the tip-sample distance in a scanning tunneling microscope (STM) may be reduced to provide stable operation at the force-equilibrium distance of 0.20 to 0.25 nm [10]. We assume that reversing the polarity of the applied voltage would give a tunneling current having approximately the same magnitude but opposite sign as is generally seen with an STM. The electrons would tunnel between the end of the tip electrode which may have a radius as small as 5 nm [11],[12] or even be a single atom [13],[14]. Because the current is directed by the intense normal electric field, that is typically greater than 1 V/nm, we assume that the flow of electrons would be normal to the anode and focused to a sub-nm spot size.

In previous analyses of the barrier traversal time for quantum tunneling [3] the author modeled quantum tunneling as the result of fluctuations in the energy of a particle that have minimum action to obtain Eq. (1) which is consistent with the uncertainty principle [15].

$$\int_{-\frac{\Delta T}{2}}^{\frac{\Delta T}{2}} \Delta E(t)\,dt \approx \frac{\hbar}{2} \qquad (1)$$

This model suggests that the most probable end points for an electron tunneling between the tip and the anode are on a normal straight line which would have minimum length to focus and collimate the electrons as they pass through wire extensions of the anode and cathode Thus, this may enable coherent propagation over a distance of up to 68.2 nm in beryllium. Single-crystal metals could be used to limit the scattering at grain boundaries. The path may also be lengthened by flaring the wires to larger diameters to further limit the interaction of the electrons with the surface of the metal.

These effects must not be confused with superconductivity because they do not require Cooper pairing in which the Anderson criterion would set a lower limit for the wire radius [16],[17].

We present two examples for nanoscale circuits with tunneling junctions. The first is heuristic because the battery and the loop structure would interfere with coherent transmission of the wavefunction. The second may be more practical because it has only a tunneling junction at the feed-point of a dipole antenna on a short straight line.

## IV. ANALYSIS OF RELEVANT WORK BY TIEN AND GORDON AS BACKGROUND

We begin with a brief description and analysis of the pioneering work by Tien and Gordon to place our effort on laser-assisted quantum tunneling in context. In 1963 Tien and Gordon [18] published the first analytical solution of the time-dependent Schrödinger equation for quantum tunneling in a sinusoidally-modulated uniform electric field. Google Scholar lists 1,007 publications that refer to this paper in studies of quantum tunneling with a variety of nanostructures.

Since Tien and Gordon did not present a circuit model, we use the one shown in Fig. 1. An ideal DC voltage source and an ideal AC voltage source are connected in series with a tunneling junction. The work function is added as two series elements in this circuit model. The planar structure that was implicit in their tunneling junction is shown by the extended dashed lines. The letters "A" and "B" denote the two sides of the model for reference. The coordinate x is zero at the surface of the cathode and equals "a" at the surface of the anode.

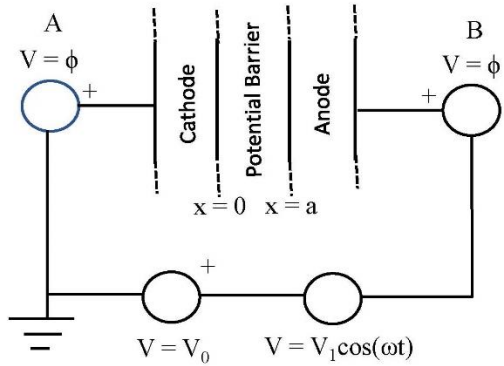

Fig. 1. Circuit model for the dynamic solution by Tien and Gordon [18].

**Formulation of the work by Tien and Gordon where we use Airy functions:**

Tien and Gordon [18] began by considering quantum tunneling between two ideal metal electrodes when there is only an applied DC electric field to cause the wavefunction $\psi_0(x,y,z)$ as shown in Eq. (2). Then they used the Transfer Hamiltonian method to derive Eq. (3) as the wavefunction that is changed by superimposing the sinusoidal potential of $V_1\cos(\omega t)$ on the DC potential of the anode.

$$\Psi_0 = \psi_0(x, y, z) e^{-i\frac{Et}{\hbar}} \qquad (2)$$

$$\Psi(x, y, z, t) = \Psi_0 \sum_{n=-\infty}^{n=+\infty} J_n\left(\frac{eV_1}{\hbar\omega}\right) e^{-in\omega t} \qquad (3)$$

In Section 1 of the Appendix we show that the static solution for the wavefunction within the potential barrier is given by Eq. (4) where Ai and Bi are Airy functions. The equations for the real constants A and B and the complex coefficients $C_1$ and $C_2$, are derived in that section. Using Eq. (5) with the identity that is derived in Section 2 of the Appendix we obtain Eq. (6).

$$\psi(x) = C_1 Ai\left(\frac{B-x}{A}\right) + C_2 Bi\left(\frac{B-x}{A}\right) \qquad (4)$$

$$\Psi(x,t) = \left[C_1 Ai\left(\frac{B-x}{A}\right) + C_2 Bi\left(\frac{B-x}{A}\right)\right] e^{-i\frac{Et}{\hbar}} \sum_{n=-\infty}^{n=+\infty} J_n\left(\frac{eV_1}{\hbar\omega}\right) e^{-in\omega t} \qquad (5)$$

$$\Psi(x,t) = \left[C_1 Ai\left(\frac{B-x}{A}\right) + C_2 Bi\left(\frac{B-x}{A}\right)\right] e^{-i\frac{Et}{\hbar} - i\frac{eV_1}{\hbar\omega}\sin(\omega t)} \qquad (6)$$

Equation (7) is the general expression for the time and spatially dependent current density within the tunneling junction. Equation (8) is the equivalent of our Eq. (3) which was derived by Tien and Gordon, as modified by using an identity we derived in Section 2 of the Appendix.

Substituting Eq. (8) into Eq. (7) gives Eq. (9) which surprisingly shows that the current density in the space between the anode and the cathode is independent of both $V_1$ and $\omega$. This is consistent with the observation by Tien and Gordon [18] that adding the time-dependent voltage does not change the spatial distribution of the wavefunction, but can only adiabatically-modify the electron energies. Thus, in the measurements that Tien and Gordon described [18], it was essential that superconductor electrodes, which changed the current because of the spectrum of the electron energies, were used to cause the measured high-frequency currents. To be clear, the analysis of Tien and Gordon does not predict that high-frequency electrical currents would be present without the superconducting metal electrodes.

$$J_X(x,t) = \frac{-ie\hbar}{2m}\left(\Psi\frac{d\Psi^*}{dx} - \Psi^*\frac{d\Psi}{dx}\right) \qquad (7)$$

$$\Psi(x,t) = \psi(x) e^{-i\frac{Et}{\hbar}} e^{-i\frac{eV_1 \sin(\omega t)}{\hbar\omega}} \qquad (8)$$

$$J_X(x) = \frac{-ie\hbar}{2m}\left[\psi\frac{d\psi^*}{dx} - \psi^*\frac{d\psi}{dx}\right] \qquad (9)$$

We acknowledge that Tien and Gordon presented a valid solution for the time-dependent Schrödinger equation and they measured high-frequency currents when using superconducting electrodes. However, we question the applications in which their analysis has been used by others that do not pertain to superconductors. Furthermore, we do not consider their solution of the time-dependent Schrödinger equation to be unique.

Other analyses that are based on the transfer Hamiltonian formulation suggest that superconducting Schottky diodes have a response with much greater nonlinearity than either metal-barrier-metal (MBM) diodes or uncooled Schottky diodes [19]. Thus, the superconducting diodes, which were first analyzed by Tien and Gordon, may be more suitable for applications as microwave mixers.

**Derivation of the barrier potential that is implicit in the analysis by Tien and Gordon:**

Tien and Gordon [18] showed that the wavefunction is given by Eq. (10) where the static solution for $V_1 = 0$ is given in Eq. (11).

$$\Psi(x,y,z,t) = \Psi_0 \sum_{n=-\infty}^{n=+\infty} J_n\left(\frac{eV_1}{\hbar\omega}\right) e^{-in\omega t} \qquad (10)$$

$$\Psi_0 = \psi(x,y,z) e^{-i\frac{Et}{\hbar}} \qquad (11)$$

Equation (12) is an identity that is derived in Section 2 of the Appendix and was confirmed by combining two equations labeled 8.514.5 and 8.515.6 in [20]. Using this identity with Eq. (10) gives Eq. (13) as the wavefunction.

$$e^{-i\alpha \sin(\beta)} \equiv \sum_{n=-\infty}^{n=+\infty} J_n(\alpha) e^{-in\beta} \tag{12}$$

$$\Psi(x,y,z,t) = \Psi_0 e^{-i\frac{eV_1 \sin(\omega t)}{\hbar \omega}} \tag{13}$$

The time-dependent Schrödinger equation for the one-dimensional potential barrier V(x,t) is given by Eq. (14). Thus, Eq. (15) which is formed by rearranging Eq. (14) may be used to determine the potential barrier that corresponds to the wavefunction that was derived by Tien and Gordon.

$$\frac{\hbar^2}{2m} \frac{\partial^2 \Psi}{\partial x^2} - V(x,t)\Psi = -i\hbar \frac{\partial \Psi}{\partial t} \tag{14}$$

$$V(x,t) = \frac{i\hbar}{\Psi} \frac{\partial \Psi}{\partial t} + \frac{\hbar^2}{2m} \frac{1}{\Psi} \frac{\partial^2 \Psi}{\partial x^2} \tag{15}$$

Solving Eq. (15) gives the wavefunction from Tien and Gordon for the case of one spatial dimension which may be written as Eq. (16).

$$\Psi(x,t) = \psi(x) e^{-i\left[\frac{Et}{\hbar} + \frac{eV_1 \sin(\omega t)}{\hbar \omega}\right]} \tag{16}$$

Substituting Eq. (16) into Eq. (15) gives the following expression for the potential barrier.

$$V(x,t) = E + eV_1 \cos(\omega t) + \frac{\hbar^2}{2m\psi} \frac{d^2 \psi}{dx^2} \tag{17}$$

Section 1 of the Appendix is a derivation for the time-independent wavefunction ψ(x) in terms of Airy functions. We present this in full detail because we have not seen this presented by others. Equations (18) and (19) are equivalent to Eqs. (A1.13) and (A1.8) in Section 1 of the Appendix, and Eq. (20) is obtained by combining Eqs. (18) and (19) to eliminate the variable ξ.

$$\frac{d^2 \psi}{d\xi^2} = -\xi \psi \tag{18}$$

$$\xi = \frac{x - B}{A} \tag{19}$$

$$\frac{d^2 \psi}{dx^2} = \frac{(B-x)}{A^3} \psi \tag{20}$$

Substituting Eq. (20) into Eq. (17) gives Eq. (21). Then substituting the definition for A from Eq. (A1.11) in Section 1 of the Appendix gives Eq. (22). Substituting the definition for B in Eq. (A1.12) of Section 1 gives Eq. (23), which is simplified to obtain Eq. (24) for the potential barrier that is implicit in the solution by Tien and Gordon.

$$V(x,t) = E + eV_1 \cos(\omega t) + \frac{\hbar^2}{2m} \frac{(B-x)}{A^3} \tag{21}$$

$$V(x,t) = E + eV_1 \cos(\omega t) + (B-x)\frac{V_0}{a} \tag{22}$$

$$V(x,t) = E + eV_1 \cos(\omega t) + \left[\left(1 + \frac{\phi - E}{V_0}\right) - \frac{x}{a}\right] V_0 \tag{23}$$

$$V(x,t) = \phi + \left(1 - \frac{x}{a}\right)V_0 + eV_1 \cos(\omega t) \tag{24}$$

## V. CRITERIA FOR CONSISTENT SIMULATIONS OF NANOSCALE CIRCUITS

**Coherent propagation of the wavefunction:**
    Others have often solved the Schrödinger equation to determine the transmission of electrons through static potential barriers having two or more parts with different constant potential energies where they have assumed that there is a single incident and a single reflected wave at one end of the model and a single transmitted wave at the other end [21],[22],[23],[24],[25]. This approach has pedagogical value, but it does not consider how the incident, reflected, and transmitted waves may interact outside of the barrier. When others have modeled a scanning tunneling microscope [26] or a field emission diode [27] generally they have only considered quantum transport between the tip and the sample electrodes. We suggest that these assumptions may not be appropriate with nanoscale circuits because of the relatively long value for the mean-free path of electrons in metals [8].
    Now we consider the possibility that quantum effects may occur throughout a nanoscale circuit because the electron mean-free path is as long as 68 nm in metallic elements [8]. The effective resistance of a nanoscale wire is actually proportional to the mean free path [9] which has been considered by others [28] However, we have already addressed this issue in Part III.

**Circuit models for consistent resistors and voltage sources:**
    In quantum simulations generally others do not show a voltage source in their diagram, or characterize this source, but they only specify the value of the applied potential. Now, in allowing for the possibility of coherent transfer of the wavefunction throughout a nanoscale circuit, we should include all of the connections and other components in a model. The voltage source may be represented by a jump in the potential at a specific point or a rise in the potential over a specified length. Figure 2 shows how a linear variation of the potential may represent a voltage source or a load resistor having specific lengths, where lossless connections in which the wavefunction has only a phase change are represented by horizontal lines. The electron energy E is shown in this figure as being greater than the maximum potential to avoid quantum tunneling in this section. It is also possible to include a constant current source as a special case of the voltage source that provides the potential which is required to maintain a specified current.

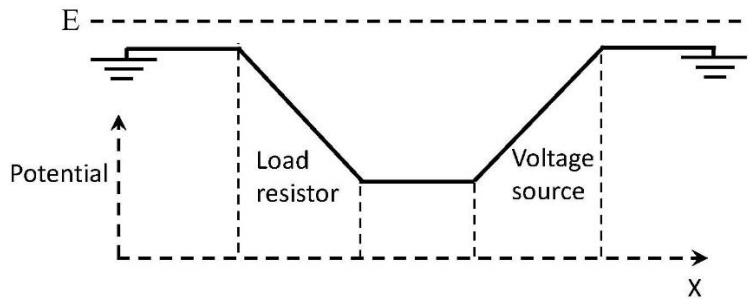

Fig. 2. Model showing a resistor, a voltage source, and lossless connectors.

    In a one-dimensional model the electrical current density in the x-direction is given by Eq. (25) as the product of the probability current density and the electron charge. The effective value

of the resistance may be estimated by dividing the voltage drop across a simulated resistor by the current, which is the product of the electrical current density and the effective cross-sectional area of the resistor. It is possible to choose a different metal, or have a different length or diameter for the wire to increase or decrease the scattering from the surface to obtain a specific resistance in a circuit. The resistor may be modeled as a lumped circuit element by having a sharp drop in the potential or by linear tapering the of the potential as shown in Fig. 2.

$$J_X(x) = \frac{-ie\hbar}{2m}\left(\psi\frac{d\psi^*}{dx} - \psi^*\frac{d\psi}{dx}\right) \qquad (25)$$

Figure 3 is a simplified closed-loop model of a nanoscale circuit having a voltage source, a load resistor, a tunneling junction and a connecting wire that is assumed to have no resistance. The two ground symbols are used to indicate that the electrical length is effectively zero between the two end-points. While only blocks, and not linear or "jump" models, are shown for the load resistor and the voltage source, it is possible to approach the problem in the following manner: First, the effective voltage may be specified as the value for the voltage source minus the drop on the load resistor. The parameters $\phi$, $U_0$, a, E, and S are specified so that the wavefunction may be determined using the Schrödinger equation. Then the electrical current density may be determined using Eq. (25). An effective cross-sectional area for the resistor may be used to estimate the actual resistance. Then the voltage drop on the load resistor may be calculated to determine the full potential of the voltage source in order to complete the solution.

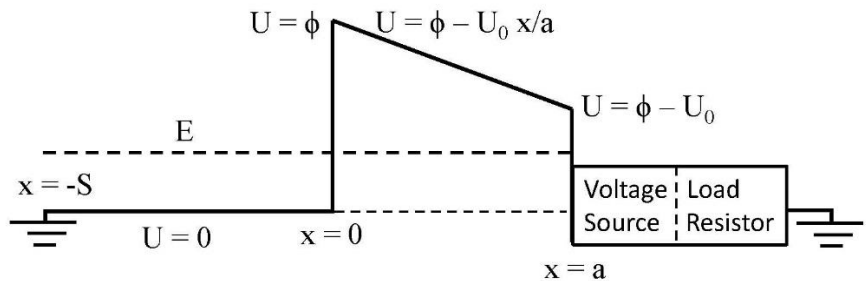

Fig. 3. Model for a closed-loop solution using a tunneling junction with a load resistor.

## VI. EXAMPLE 1: STATIC SOLUTION OF THE SCHRÖDINGER EQUATION

Figure 4 shows the potential energy in a model of a nanoscale circuit for a static problem that includes quantum tunneling in a uniform electric field. Region 3 is the battery, which has a linear variation in the potential. The "battery" could be a resistor fed by two relatively long wires connected to an external voltage source or a resistor connected to a relatively long dipole antenna for coupling to an external source. Airy functions are used to model the wavefunction in both the barrier and the battery. We follow a procedure with the Airy functions that is similar to that in the derivation in Section 1 of the Appendix. For clarity, the symbols "U" and "V" are used separately to denote the potential energy and the voltage.

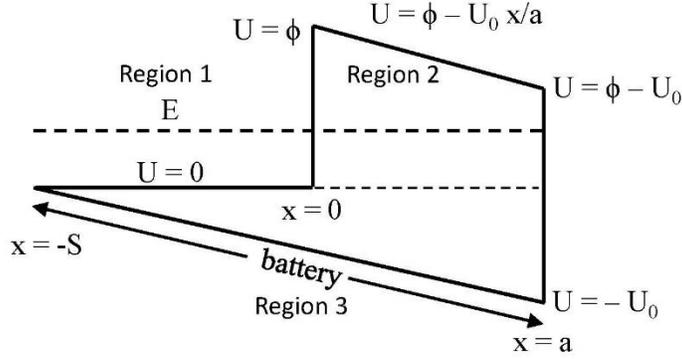

Fig. 4. Potential energy for a model of quantum tunneling in a static potential barrier.

**Determine expressions for the wavefunctions:**

In Region 3, which represents the battery where -S < x < a, the potential is given by Eq. (26). Substituting this into the Schrödinger equation given in Eq. (27) results in Eq. (28).

$$U_3(x) = -U_0 \frac{(S+x)}{(S+a)} \tag{26}$$

$$\frac{\hbar^2}{2m}\frac{d^2\psi_3}{dx^2} + [E - U(x)]\psi_3 = 0 \tag{27}$$

$$\frac{\hbar^2}{2m}\frac{d^2\psi_3}{dx^2} + \left[E + \frac{U_0 S}{(S+a)} + \frac{U_0 x}{(S+a)}\right]\psi_3 = 0 \tag{28}$$

A change of variables shown in Eq. (29) is used with Eq. (28) to obtain Eq. (30) where the coefficients $A_3$ and $B_3$ have units of meters. Then Eq. (30) is rearranged to obtain Eq. (31).

$$x = A_3 \xi + B_3 \tag{29}$$

$$\frac{\hbar^2}{2mA_3^2}\frac{d^2\psi_3}{d\xi^2} + \left[E + \frac{U_0 S}{(S+a)} + \frac{U_0(A_3\xi + B_3)}{(S+a)}\right]\psi_3 = 0 \tag{30}$$

$$\frac{d^2\psi_3}{d\xi^2} + \frac{2mA_3^2}{\hbar^2}\left[E + \frac{U_0(S+B_3)}{(S+a)}\right]\psi_3 + \frac{2mA_3^3 U_0}{\hbar^2(S+a)}\xi\psi_3 = 0 \tag{31}$$

Parameter $A_3$ is chosen so that the coefficient of $\xi\psi_3$ in Eq. (31) is unity and parameter $B_3$ is chosen so that the quantity that is in brackets in Eq. (31) is zero. Thus, parameters $A_3$ and $B_3$ are given in Eqs. (32) and (33), and Eq. (30) is simplified to give Eq. (34). Note that $A_3$ is greater than zero and $B_3$ is negative.

$$A_3 = \left[\frac{\hbar^2(S+a)}{2mU_0}\right]^{\frac{1}{3}} \tag{32}$$

$$B_3 = -(S+a)\frac{E}{U_0} - S \tag{33}$$

$$\frac{d^2\psi_3}{d\xi^2} + \xi\psi_3 = 0 \tag{34}$$

The solution of Eq. (34) is given in Eq. (35) where Ai and Bi are Airy functions [29]. Next Eq. (29) is used with Eq. (35) to obtain Eq. (36) so that x is again the independent variable. In Region 3 the sign of the argument for the Airy functions is negative to give a quasi-sinusoidal behavior because the energy is greater than the potential. However, in Region 2, depending on the value for the energy E, there may be quantum tunneling in none, part, or all of the length of the barrier. The solutions with the Airy functions make it simpler to implement the boundary conditions than seen in the transition between real and imaginary exponentials in Region 1.

Taking the derivative of Eq. (36) gives Eq. (37) for the derivative of the wavefunction in Region 3.

$$\psi_3(\xi) = C_5 Ai(-\xi) + C_6 Bi(-\xi) \tag{35}$$

$$\psi_3(x) = C_5 Ai\left(\frac{B_3 - x}{A_3}\right) + C_6 Bi\left(\frac{B_3 - x}{A_3}\right) \tag{36}$$

$$\frac{d\psi_3}{dx} = -\frac{C_5}{A_3} Ai'\left(\frac{B_3 - x}{A_3}\right) - \frac{C_6}{A_3} Bi'\left(\frac{B_3 - x}{A_3}\right) \tag{37}$$

The wavefunction and its derivative in Region 1, where $-S < x < 0$, are given by Eqs. (38), (39), and (40).

$$\psi_1(x) = C_1 e^{-ik_1 x} + C_2 e^{ik_1 x} \tag{38}$$

$$k_1 = \frac{\sqrt{2mE}}{\hbar} \tag{39}$$

$$\frac{d\psi_1}{dx} = -ik_1 C_1 e^{-ik_1 x} + ik_1 C_2 e^{ik_1 x} \tag{40}$$

In Section 3 of the Appendix it is shown that the current density is independent of x in Region 1.

In Region 2, where $0 < x < a$, the potential is given by Eq. (41) so the Schrödinger equation is given by Eq. (42).

$$U_2 = \phi - U_0 \frac{x}{a} \tag{41}$$

$$\frac{\hbar^2}{2m} \frac{d^2 \psi_2}{dx^2} + \left[E - \phi + U_0 \frac{x}{a}\right] \psi_2 = 0 \tag{42}$$

The change of variables in Eq. (43) is used to obtain Eq. (44), which is rearranged to form Eq. (45).

$$x = A_2 \xi + B_2 \tag{43}$$

$$\frac{\hbar^2}{2mA_2^2} \frac{d^2 \psi_2}{d\xi^2} + \left[E - \phi + \frac{B_2}{a} U_0 + A_2 \frac{\xi}{a} U_0\right] \psi_2 = 0 \tag{44}$$

$$\frac{\hbar^2}{2mA_2^2} \frac{d^2 \psi_2}{d\xi^2} + \left[E - \phi + B_2 \frac{U_0}{a}\right] \psi_2 + A_2 \frac{\xi}{a} U_0 \psi_2 = 0 \tag{45}$$

Parameter $B_2$, which is positive because E is less than $\phi$, is chosen as shown in Eq. (46) to negate the second term in Eq. (45) and obtain Eq. (47).

$$B_2 = \frac{(\phi - E)}{U_0} a \tag{46}$$

$$\frac{d^2\psi_2}{d\xi^2} + \frac{2mU_0 A_2^3}{\hbar^2 a}\xi\psi_2 = 0 \quad (47)$$

Parameter $A_2$, which is also positive, is chosen as shown in Eq. (48) to simplify the second term in Eq. (47) to obtain Eq. (49).

$$A_2 = \left(\frac{\hbar^2 a}{2mU_0}\right)^{\frac{1}{3}} \quad (48)$$

$$\frac{d^2\psi_2}{d\xi^2} + \xi\psi_2 = 0 \quad (49)$$

The wavefunction for Region 2 is given by Eq. (50), which is the solution of Eq. (49), and the derivative of this wavefunction is shown in Eq. (51).

$$\psi_2(x) = C_3 Ai\left(\frac{B_2 - x}{A_2}\right) + C_4 Bi\left(\frac{B_2 - x}{A_2}\right) \quad (50)$$

$$\frac{d\psi_2}{dx} = -\frac{C_3}{A_2} Ai'\left(\frac{B_2 - x}{A_2}\right) - \frac{C_4}{A_2} Bi'\left(\frac{B_2 - x}{A_2}\right) \quad (51)$$

Section 4 of the Appendix uses Eq. (50) to interpret the wavefunction in Region 2 in terms of forward and reflected waves within the barrier.

**Apply the boundary conditions:**

When x equals 0 the argument of the Airy functions in Eq. (50) is given by Eq. (52), and at x = a, the argument is given by Eq. (53).

$$Arg_1 = \frac{(\phi - E)a}{A_2 U_0} > 0 \quad (52)$$

$$Arg_2 = \frac{(\phi - E)a - U_0 a}{A_2 U_0} = \frac{(\phi - U_0 - E)a}{A_2 U_0} \quad (53)$$

It may be seen in Fig. 4 that, for $E < \phi - U_0$, the electron will tunnel through the full length of the barrier from x = 0 to x = a. The argument of the Airy functions in Eq. (50) is decreased during this transit but is smaller, while non-zero, at x = a.

Next the two boundary conditions, that the wavefunction and its derivative are continuous, are applied at the three boundaries. First Eqs. (54) and (55) are obtained at x = 0 between Region 1 and Region 2.

$$C_1 + C_2 - Ai\left(\frac{B_2}{A_2}\right)C_3 - Bi\left(\frac{B_2}{A_2}\right)C_4 = 0 \quad (54)$$

$$ik_1 A_2 C_1 - ik_1 A_2 C_2 - Ai'\left(\frac{B_2}{A_2}\right)C_3 - Bi'\left(\frac{B_2}{A_2}\right)C_4 = 0 \quad (55)$$

Applying the boundary conditions at x = a, between Region 2 and Region 3, gives Eqs. (56) and (57).

$$Ai\left(\frac{B_2 - a}{A_2}\right)C_3 + Bi\left(\frac{B_2 - a}{A_2}\right)C_4 - Ai\left(\frac{B_3 - a}{A_3}\right)C_5 - Bi\left(\frac{B_3 - a}{A_3}\right)C_6 = 0 \quad (56)$$

$$Ai'\left(\frac{B_2-a}{A_2}\right)C_3 + Bi'\left(\frac{B_2-a}{A_2}\right)C_4 - \frac{A_2}{A_3}Ai'\left(\frac{B_3-a}{A_3}\right)C_5 - \frac{A_2}{A_3}Bi'\left(\frac{B_3-a}{A_3}\right)C_6 = 0 \quad (57)$$

Applying the boundary conditions at x = -S, between Region 3 and Region 1, gives Eqs. (58) and (59).

$$e^{ik_1S}C_1 + e^{-ik_1S}C_2 - Ai\left(\frac{B_3+S}{A_3}\right)C_5 - Bi\left(\frac{B_3+S}{A_3}\right)C_6 = 0 \quad (58)$$

$$ik_1A_3 e^{ik_1S}C_1 - ik_1A_3 e^{-ik_1S}C_2 - Ai'\left(\frac{B_3+S}{A_3}\right)C_5 - Bi'\left(\frac{B_3+S}{A_3}\right)C_6 = 0 \quad (59)$$

**Define the matrix elements:**

Equations (54) through (59) form a system of six simultaneous homogeneous equations in the six unknown complex coefficients $C_1$ through $C_6$. We define the elements of the matrix for this system as $M_{IJ}$ where the indices I and J are the row and column numbers which each run from 1 through 6. The equation on each row has two matrix elements which are zero, and the 24 non-zero matrix elements are given in equations (60) to (83):

$$M_{11} = 1 \quad (60)$$

$$M_{12} = 1 \quad (61)$$

$$M_{13} = -Ai\left(\frac{B_2}{A_2}\right) \quad (62)$$

$$M_{14} = -Bi\left(\frac{B_2}{A_2}\right) \quad (63)$$

$$M_{21} = ik_1 A_2 \quad (64)$$

$$M_{22} = -ik_1 A_2 \quad (65)$$

$$M_{23} = -Ai'\left(\frac{B_2}{A_2}\right) \quad (66)$$

$$M_{24} = -Bi'\left(\frac{B_2}{A_2}\right) \quad (67)$$

$$M_{33} = Ai\left(\frac{B_2-a}{A_2}\right) \quad (68)$$

$$M_{34} = Bi\left(\frac{B_2-a}{A_2}\right) \quad (69)$$

$$M_{35} = -Ai\left(\frac{B_3-a}{A_3}\right) \quad (70)$$

$$M_{36} = -Bi\left(\frac{B_3-a}{A_3}\right) \quad (71)$$

$$M_{43} = Ai'\left(\frac{B_2-a}{A_2}\right) \quad (72)$$

$$M_{44} = Bi'\left(\frac{B_2 - a}{A_2}\right) \tag{73}$$

$$M_{45} = -\frac{A_2}{A_3} Ai'\left(\frac{B_3 - a}{A_3}\right) \tag{74}$$

$$M_{46} = -\frac{A_2}{A_3} Bi'\left(\frac{B_3 - a}{A_3}\right) \tag{75}$$

$$M_{51} = e^{ik_1 S} \tag{76}$$

$$M_{52} = e^{-ik_1 S} \tag{77}$$

$$M_{55} = -Ai\left(\frac{B_3 + S}{A_3}\right) \tag{78}$$

$$M_{56} = -Bi\left(\frac{B_3 + S}{A_3}\right) \tag{79}$$

$$M_{61} = ik_1 A_3 e^{ik_1 S} \tag{80}$$

$$M_{62} = -ik_1 A_3 e^{-ik_1 S} \tag{81}$$

$$M_{65} = -Ai'\left(\frac{B_3 + S}{A_3}\right) \tag{82}$$

$$M_{66} = -Bi'\left(\frac{B_3 + S}{A_3}\right) \tag{83}$$

Using the above notation, the system of Eqs. (54) through (59) may be written as follows where we implement the values of $M_{11} = 1$ and $M_{12} = 1$:

$$C_1 + C_2 + M_{13}C_3 + M_{14}C_4 = 0 \tag{84}$$

$$M_{21}C_1 + M_{22}C_2 + M_{23}C_3 + M_{24}C_4 = 0 \tag{85}$$

$$M_{33}C_3 + M_{34}C_4 + M_{35}C_5 + M_{36}C_6 = 0 \tag{86}$$

$$M_{43}C_3 + M_{44}C_4 + M_{45}C_5 + M_{46}C_6 = 0 \tag{87}$$

$$M_{51}C_1 + M_{52}C_2 + M_{55}C_5 + M_{56}C_6 = 0 \tag{88}$$

$$M_{61}C_1 + M_{62}C_2 + M_{65}C_5 + M_{66}C_6 = 0 \tag{89}$$

## VII. DETERMINE VALID SETS FOR THE PARAMETERS IN EXAMPLE 1.

The system of Equations (84) through (89) is homogeneous. Thus, in order to have a unique non-trivial solution the coefficients must be chosen so that the determinant of the matrix is zero. Others have previously extended the Saurus "short-cut" for calculating the determinant of 3 by 3 matrices to enable applications with 4 x 4 matrices [30]. We acknowledge that in two preliminary presentations of our analysis the Saurus method was incorrectly applied to the present set of 6 by 6 complex matrices [31],[32]. Now we have used Cramer's rule as an accepted method to evaluate the determinant.

The expanded determinant has 80 non-zero terms and all except 8 contain one or more members of the group $M_{51}$, $M_{52}$, $M_{61}$, and $M_{62}$ which are complex. Thus, it is not a simple matter to simplify the solution by separating the real and imaginary parts of the 80 terms. It is possible to force $M_{51}$ and $M_{52}$ to be real by requiring that $k_1 S$ is equal to $n\pi$, or to make $M_{61}$ and $M_{62}$ real by requiring that $k_1 S$ is equal to $n\pi/2$. However, these two conditions, which would permit separately setting the real or imaginary parts of the determinant to zero, cannot be satisfied simultaneously.

When values are specified for any 4 of the parameters in the set $\{\phi, a, U_0, E, S\}$ test values for the remaining parameter may be used to determine the matrix elements $M_{IJ}$ with Eqs. (60) to (83). Then the remaining parameter may be changed to minimize the magnitude of the determinant that is defined by Eqs. (54) to (59), or equivalently Eqs. (84) to (89). In doing this the potential energy $U_0$ is determined from the specified value of the applied voltage $V_0$. We find that there are sharply-defined minima for the determinant. At each minimum for the magnitude of the determinant we use the lowest calculated value as the zero and determine the uncertainty for this value by reducing the spacing between the closest points to each side of the minimum.

Figures 5 through 10 were prepared using points with $10^5$ calculated values evenly spaced on the linear range of the abscissa. For example, the spacing between consecutive data points was $10^{-4}$ nm in Fig. 5, and $2 \times 10^{-5}$ nm in Fig. 8. Consistency, but finer detail, is seen when comparing the plots made using different values for the range, such as in figures 5 and 6.

Figure 5 shows the logarithm of the magnitude of the determinant as a function of a, which is the length of the tunneling junction. The base of the dip at each resonance is zero which does not show in this figure, even though the length of the tunneling junction has a resolution of $9.999 \times 10^{-5}$ nm, because the resonances are extremely sharp. The zeros are evenly-spaced with values of a equal to 0.7032, 1.300, 1.9043, 2.5114, 3.1199, 3.7292, 0.43389, 5.5591, 6.1694, 6.7799, 8.0009, 8.6116, 9.2222, and 9.8328 nm. The best fit of these data as a straight line has a slope of 0.609363, an intercept of 0.0780425, an r-value of 0.999998, a p-value of $1.12425 \times 10^{-39}$, and a standard error of 0.00030076.

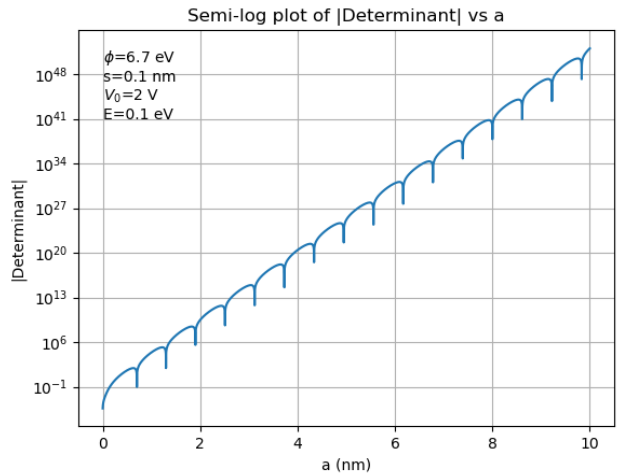

Fig. 5. Magnitude of the determinant as a function of the length of the tunneling junction where the work function is 6.7 eV, the applied potential is 2 V, the electron energy is 0.1 eV, and the pre-barrier length is 0.1 nm

Figure 6 is a linear plot showing a closeup of the data from Fig. 5 near the second resonance at which a = 1.300 nm. The need to obtain fine resolution to show the extremely sharp nature of the resonances is seen by comparing these two figures.

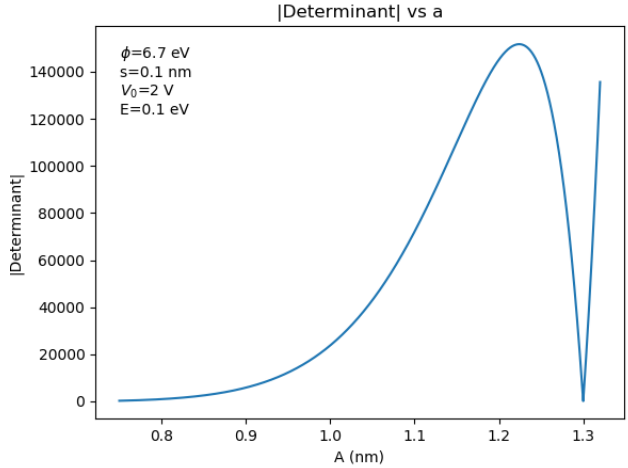

Fig. 6. Magnitude of the determinant as a function of the length of the tunneling junction where the work function is 6.7 eV, the applied potential is 2 V, the electron energy is 0.1 eV, and the pre-barrier length is 0.1 nm. The second resonance is at a = 1.300 nm.

Figure 7 is an extreme closeup showing the data from Fig. 5 to show the residuals of the determinant near the first resonance where a is equal to 0.07032 nm. The spacing between consecutive points is $10^{-4}$ nm, as in Figures 5 and 6, but now each data point is seen because of the finer resolution. It is surprising that there is a lack of symmetry in this figure.

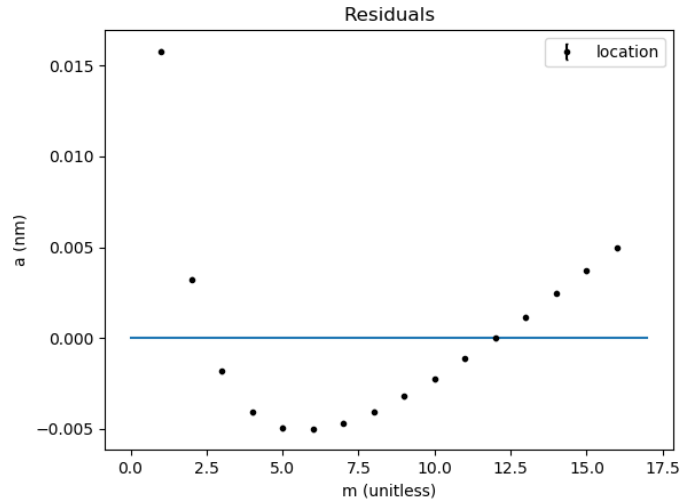

Fig. 7. Plot of 16 data points for the determinant near the first resonance in Fig. 5 where a ≈ 0.07032 nm. The spacing between points on the abscissa is $10^{-4}$ nm as in Figs. 5 and 6, and the width of the zero-crossing is $10^{-3}$ nm.

Figure 8 shows the magnitude of the determinant as a function of S, the length of the pre-barrier region. The spacing between consecutive points on the abscissa is 2.0 x$10^{-5}$ nm. The first two zeros of the determinant occur when S equals 1.05085 and 1.71210 nm.

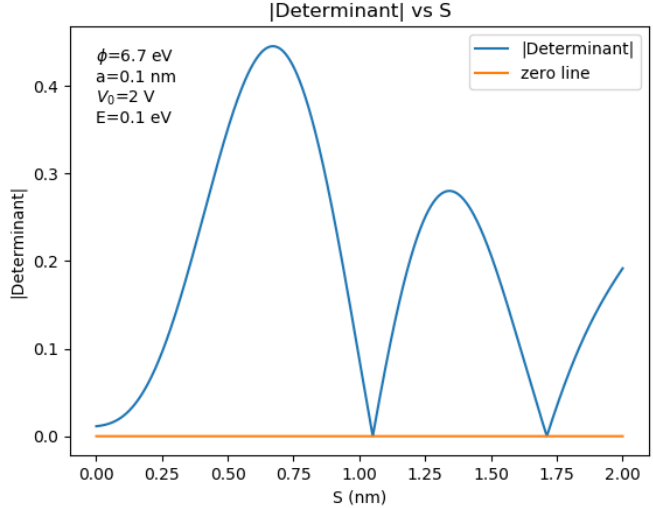

Fig. 8. Magnitude of the determinant as a function of the length of the pre-barrier region where the length of the tunneling junction is 0.1 nm. The work function is 6.7 eV, the potential is 2.0 V, and the electron energy is 0.1 eV,

Figure 9 shows the magnitude of the determinant as a function of the applied potential $V_0$. The length of the tunneling junction is 0.1 nm. The work function is 6.7 eV and the electron energy is 0.1 eV. The length of the pre-barrier region is 0.5 nm. The first resonance is at a potential of 9.30019 V. Notice that the multiple zeros for the determinant are not evenly spaced.

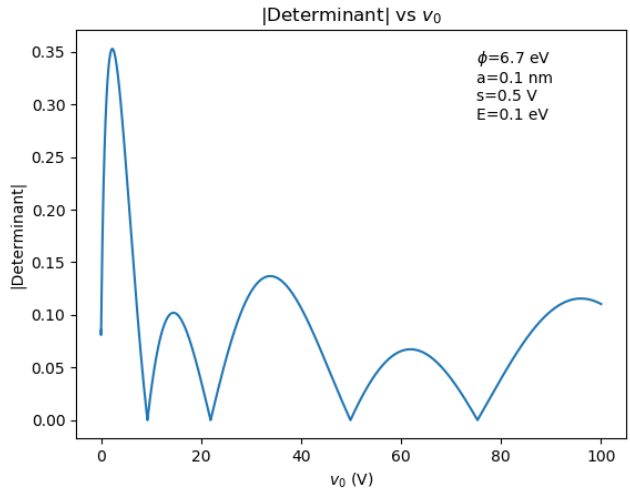

Fig. 9 shows the magnitude of the determinant as a function of the applied potential where the length of the tunneling junction is 0.1 nm, the work function is 6.7 eV and the electron energy is 0.1 eV. The length of the pre-barrier region is 0.5 nm and the first zero is at a potential of 9.30019 V.

Figure 10 shows the magnitude of the determinant as a function of E, which is the energy of the electrons. The length of the tunneling junction is 0.1 nm and the work function is 6.7 eV. The applied potential is 2 V and the pre-barrier length is 0.1 nm. We are surprised to see that there is no resonance in this figure.

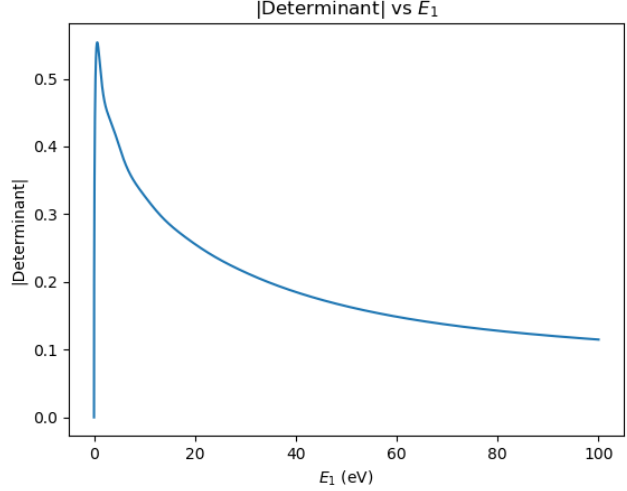

Fig. 10. Magnitude of the determinant as a function of the energy of the electrons where the length of the tunneling junction is 0.1 nm and the work function is 6.7 eV. The applied potential is 2 V and the pre-barrier length is 0.1 nm.

**VIII. DETERMINE THE CORRESPONDING NORMALIZED COEFFICIENTS.**

Equations (84) through (89) may be normalized by dividing all of the terms by the coefficient $C_1$ and defining the normalized coefficients $C_{IN} \equiv C_I/C_1$ for I = 2 through 6 to obtain Eqs. (90) to (95).

$$1 + C_{2N} + M_{13}C_{3N} + M_{14}C_{4N} = 0 \qquad (90)$$
$$M_{21} + M_{22}C_{2N} + M_{23}C_{3N} + M_{24}C_{4N} = 0 \qquad (91)$$
$$M_{33}C_{3N} + M_{34}C_{4N} + M_{35}C_{5N} + M_{36}C_{6N} = 0 \qquad (92)$$
$$M_{43}C_{3N} + M_{44}C_{4N} + M_{45}C_{5N} + M_{46}C_{6N} = 0 \qquad (93)$$
$$M_{51} + M_{52}C_{2N} + M_{55}C_{5N} + M_{56}C_{6N} = 0 \qquad (94)$$
$$M_{61} + M_{62}C_{2N} + M_{65}C_{5N} + M_{66}C_{6N} = 0 \qquad (95)$$

Moving the terms that are without a normalized coefficient to the right-hand side (RHS) in each of these 6 equations gives the following system of equations which we solve to determine the values of the normalized coefficients. Any one of these six equations may be derived from the other five so one equation may be deleted to obtain a determined system of 5 equations that is solved to determine the 5 normalized coefficients.

$$C_{2N} + M_{13}C_{3N} + M_{14}C_{4N} = -1 \qquad (96)$$
$$M_{22}C_{2N} + M_{23}C_{3N} + M_{24}C_{4N} = -M_{21} \qquad (97)$$
$$M_{33}C_{3N} + M_{34}C_{4N} + M_{35}C_{5N} + M_{36}C_{6N} = 0 \qquad (98)$$
$$M_{43}C_{3N} + M_{44}C_{4N} + M_{45}C_{5N} + M_{46}C_{6N} = 0 \qquad (99)$$
$$M_{52}C_{2N} + M_{55}C_{5N} + M_{56}C_{6N} = -M_{51} \qquad (100)$$
$$M_{62}C_{2N} + M_{65}C_{5N} + M_{66}C_{6N} = -M_{61} \qquad (101)$$

Using the definitions for the matrix elements in Eqs. (60) to (83), the system of equations (96) through (101) may also be written as Eqs. (102) to (107).

$$C_{2N} - Ai\left(\frac{B_2}{A_2}\right)C_{3N} - Bi\left(\frac{B_2}{A_2}\right)C_{4N} = -1 \quad (102)$$

$$ik_1 A_2 C_{2N} + Ai'\left(\frac{B_2}{A_2}\right)C_{3N} + Bi'\left(\frac{B_2}{A_2}\right)C_{4N} = ik_1 A_2 \quad (103)$$

$$Ai\left(\frac{B_2 - a}{A_2}\right)C_{3N} + Bi\left(\frac{B_2 - a}{A_2}\right)C_{4N} - Ai\left(\frac{B_3 - a}{A_3}\right)C_{5N} - Bi\left(\frac{B_3 - a}{A_3}\right)C_{6N} = 0 \quad (104)$$

$$Ai'\left(\frac{B_2 - a}{A_2}\right)C_{3N} + Bi'\left(\frac{B_2 - a}{A_2}\right)C_{4N} - \frac{A_2}{A_3}Ai'\left(\frac{B_3 - a}{A_3}\right)C_{5N} - \frac{A_2}{A_3}Bi'\left(\frac{B_3 - a}{A_3}\right)C_{6N} = 0 \quad (105)$$

$$e^{-ik_1 S}C_{2N} - Ai\left(\frac{B_3 + S}{A_3}\right)C_{5N} - Bi\left(\frac{B_3 + S}{A_3}\right)C_{6N} = -e^{ik_1 S} \quad (106)$$

$$ik_1 A_3 e^{-ik_1 S}C_{2N} + Ai'\left(\frac{B_3 + S}{A_3}\right)C_{5N} + Bi'\left(\frac{B_3 + S}{A_3}\right)C_{6N} = ik_1 A_3 e^{ik_1 S} \quad (107)$$

A procedure for determining the set of 5 normalized coefficients is illustrated in Section 5 of the Appendix.

## IX. EXAMPLE 2: QUASISTATIC SOLUTION OF THE SCHRÖDINGER EQUATION

Figure 11 shows the potential energy in a model of a nanoscale circuit in which a tunneling junction is between two electrically-short monopole antennas, all on a straight line, in a uniform electric field $E_x$. Calculations for a sequence of different values for $E_x$ could be used as a quasistatic approximation for the response of this model to a time-dependent field. It is assumed that the potential is independent of the coordinate x on each of the two monopoles because they are much shorter than the wavelength but the wavefunction is zero at the outer ends of the two monopoles. In a device based on this model the diameters of the two wire monopoles could be increased or tapered outward to reduce the loss by further decreasing the interaction of the electrons with the outer surfaces of these two wires.

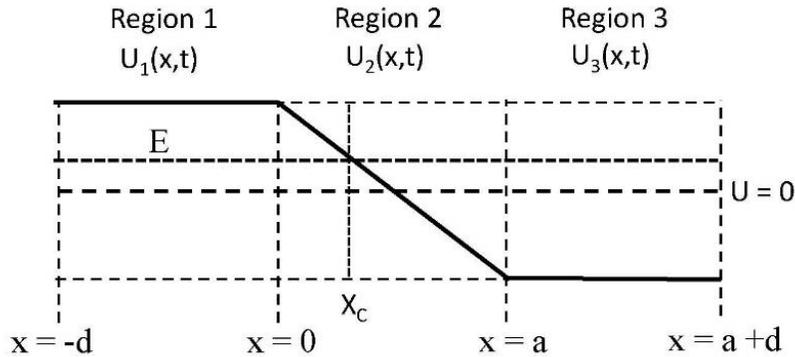

Fig. 11. Potential energy U for a model with quasistatic excitation of a tunneling junction shown at a time when $E_x < 0$.

The potential energy for an electron with charge -e in the three regions is given in Eqs. (108), (109), and (110). There is a uniform electric field in Region 2 and no electric field in regions 1 and 3. Notice that for an electron, because of its negative charge, the dominant transport would be from right to left at times when $E_x$ is negative as shown in Fig. 11.

The potential energy, which is a function of both x and t, is given for the three regions in Eqs. (108), (109), and (110).

In Region 1, where -d < x < 0:

$$U_1 = -eE_x \frac{a}{2} \tag{108}$$

In Region 2, where 0 < x < a:

$$U_2 = -eE_x \left(\frac{a}{2} - x\right) \tag{109}$$

In Region 3, where a < x < a + d:

$$U_3 = eE_x \frac{a}{2} \tag{110}$$

In Region 2 the potential energy U intersects the energy E at $x = X_C$, which is given by Eq. (111). If E were zero this crossing would be at x = a/2 but otherwise this location depends on $X_C$ as shown in Fig. 11 when Ex is negative.

$$X_C = \frac{a}{2} + \frac{E}{eE_x} \tag{111}$$

**Algorithm to determine sets of the four parameters that are required for solutions:**

As in Example 1, a system of simultaneous equations is formed to satisfy the boundary conditions. This system is homogeneous so the determinant must be zero for a non-trivial solution. This occurs for specific values of the set of parameters {a, d, E, and $E_x$} that may be determined by using the following algorithm:

**1. Specify the value of Ex.**
Then use Eqs. (108) and (110) to calculate $U_1$ and $U_3$.

**2. Specify the value of the energy E.**
If E is greater than $U_1$ calculate $k_{1A}$ with Eq. (112) and use $\psi_{1A}$ in Eq. (113) as $\psi_1$.
However, if E is less than $U_1$ calculate $\gamma_{1B}$ with Eq. (114) and use $\psi_{1B}$ in Eq. (115) as $\psi_1$.

$$k_{1A} = \frac{1}{\hbar}\sqrt{2m(E - U_1)} \tag{112}$$

$$\psi_{1A} = A_{1A} e^{-ik_{1A}x} + B_{1A} e^{ik_{1A}x} \tag{113}$$

$$\gamma_{1B} = \frac{1}{\hbar}\sqrt{2m(U_1 - E)} \tag{114}$$

$$\psi_{1B} = C_{1B} e^{-\gamma_{1B}x} + D_{1B} e^{\gamma_{1B}x} \tag{115}$$

If E is greater than $U_3$ calculate $k_{3A}$ with Eq. (116) and use $\psi_{3A}$ in Eq. (117) as $\psi_3$.
However, if E is less than $U_3$ calculate $\gamma_{3B}$ with Eq. (118) and use $\psi_{3B}$ in Eq. (119) as $\psi_3$.

$$k_{3A} = \frac{1}{\hbar}\sqrt{2m(E - U_3)} \tag{116}$$

$$\psi_{3A} = A_{3A} e^{-ik_{3A}x} + B_{3A} e^{ik_{3A}x} \tag{117}$$

$$\gamma_{3B} = \frac{1}{\hbar}\sqrt{2m(U_3 - E)} \tag{118}$$

$$\psi_{3B} = C_{3B}e^{-\gamma_{3B}x} + D_{3B}e^{\gamma_{3B}x} \qquad (119)$$

Calculate $A_2$ and $B_2$ using Eqs. (120) and (121) so that we use Eq. (122) to determine $\psi_2$. The derivation for Eqs. (120), (121) and (122) is in Section 6 of the Appendix.

$$A_2 = -\left(\frac{\hbar^2}{2meE_x}\right)^{\frac{1}{3}} \qquad (120)$$

$$B_2 = \frac{a}{2} + \frac{E}{eE_x(t)} \qquad (121)$$

$$\psi_2 = F_2 A_i\left(\frac{B_2 - x}{A_2}\right) + G_2 B_i\left(\frac{B_2 - x}{A_2}\right) \qquad (122)$$

**3. Apply the boundary conditions to determine the matrix elements.**

For all values of the energy E other than zero different values of $E_x$ cause either (1) quantum tunneling and classical propagation, (2) only quantum tunneling, or (3) only classical propagation in this model. In Section 7 of the Appendix the boundary conditions are applied to determine expressions for the matrix elements in each of the 4 possible cases:

Case 1: $E > U_1$ and $U_3$ are Eqs. (A7.1) through (A7.6).
Case 2: $U_1 < E < U_3$ are Eqs. (A7.7) through (A7.12).
Case 3: $U_3 < E < U_1$ are Eqs. (A7.13) through (A7.18).
Case 4: $E < U_1$ and $U_3$ are Eqs. (A7.19) through (A7.24).

**4. Obtain the 4 determinants and set them to zero to determine unique sets for parameters.**

In each of the 4 cases these groups of 4 simultaneous homogeneous equations are used to form a matrix in which the determinant is set to zero to obtain non-trivial solutions for the set of parameters as was done in Part VII for Example 1.

**5. Determine the corresponding normalized coefficients for each mode.**

Follow the procedures that were already applied in Part VIII for Example 1 to determine the normalized coefficients for each mode. This may be done explicitly following the procedure used in Section 5 of the Appendix for Example 1, or simply by solving the corresponding matrix equation.

## X. EXTENSION TO APPLICATIONS WITH TIME-DEPENDENT POTENTIALS

**Applications to continuous wave (CW) lasers:**

In Part III we noted that the mechanism studied by Tien and Gordon with a CW laser [18] requires photon processes with superconducting electrodes to obtain a time-dependent tunneling current, and also concluded that their solution of the time-dependent Schrödinger equation is not unique. A separate exact solution of the time-dependent Schrödinger equation, which also requires photon processes, was presented more recently by Zhang and Lau [33].

Others have studied different types of laser-assisted assisted tunneling by both analyses [18],[33].[34],[35] and measurements [36],[37],[38]. For example, others generated a 435 MHz beat signal by optical heterodyning two modes of a He-Ne laser focused on the tip-sample junction of an STM [39] and generated microwave signals tunable from 2 to 13 GHz by focusing two infrared lasers on an yttrium-iron-garnet film at the tunneling junction of an STM [40]. A summary of related work by others was published by Grafström [41].

In our first (unpublished) measurements of quantum tunneling with time-dependent applied potentials we connected the primary windings of three audio-frequency transformers in series with a sealed vacuum field emission tube and an ungrounded DC high-voltage power supply. Two audio-frequency oscillators and an oscilloscope were connected in series with the secondary windings of the three audio-frequency transformers. Photon processes were not considered when interpreting the measurements made using the oscilloscope because the photon energy is on the order of $10^{-12}$ eV at the frequencies which were used.

The standard Fowler-Nordheim model, and more recent revisions of this model [42] are often used to determine the current density as a function of the applied electric field in field emission. The expression for the standard Fowler-Nordheim model is shown as Eq. (123) where "J" denotes the current density, "F" denotes the applied electric field, and A and B are constants that depend on the parameters for a given device.

$$J = AF^2 e^{-\frac{B}{F}} \tag{123}$$

The constant F is proportional to the total applied potential that was used in our measurements. The voltages from the two audio-frequency oscillators were less than 1 percent of the applied DC voltage to cause a small but measurable perturbation in the total current. When using Eq. (123) to predict the relative values for the currents that we measured at the harmonics and mixer frequencies there was reasonable agreement between our measurements and the standard Fowler-Nordheim model.

**Applications to mode-locked lasers:**

More recently we have generated microwave frequency combs which have hundreds of harmonics at integer multiples of the pulse-repetition frequency (74.254 MHz) of a mode-locked ultrafast laser. This was done by focusing the laser on the tunneling junction of a scanning tunneling microscope [5]. A gold sample electrode ($\phi$ = 5.5 eV) was used with a tungsten tip ($\phi$ = 4.5 eV). The Ti: Sapphire laser has a spectrum extending from 650 to 1180 nm (1.91 eV to 1.05 eV) with a center wavelength of 800 nm (1.55 eV) for an energy that is much lower than the work functions for the sample and tip electrodes. Thus, photon processes are unlikely, which was confirmed by an analysis showing that these measurements are consistent with analysis using a quasistatic approximation [43].

The harmonics of the microwave frequency comb were measured with a spectrum analyzer that was connected to a Bias-T inserted into the sample circuit of the STM. Figure 12 is an equivalent circuit that we have used to understand the effects of this circuit on our measurements of the microwave harmonics. When using the STM (UHV700, from RHK Technology) with the mode-locked laser we see that the power which is measured at the harmonics varies inversely as the square of the frequency and the time constant for this roll-off is equivalent to the effect of the input impedance of 50 $\Omega$ for the spectrum analyzer shunted by the capacitance of 6.4 pF that is caused by the leads and the tunneling junction [5]. Our analysis shows that in an ideal system, where the shunting capacitance would be eliminated, there would be no significant roll-off of the measured output until a frequency of 7 THz [44].

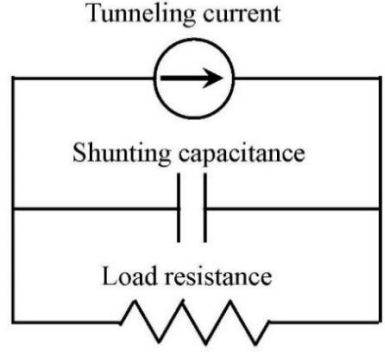

Fig. 12. Equivalent circuit measurements in laser-assisted scanning tunneling microscopy.

We use what we call a "Sequential Quasistatic Approximation" to determine the spectrum for the harmonics that are caused by a mode-locked laser. In this approach the applied potential is approximated by a sequence of values at different times and the waveform for the output current is obtained by placing the corresponding values for the current, which are calculated using our algorithms, in the same sequence. In doing so it is necessary to introduce the time-dependent phenomena that are caused by reactive elements as we have described in the previous paragraph.

Our published simulations suggest that the nonlinear current-voltage relationship of the tunneling junction causes the observed response to both audio-frequency and optical-frequency inputs because the quantum energy is well below the work functions of the electrodes in the tunneling junction [43]. Thus, a sequence of values for the electric field $E_x$ may be used with the analysis in Example 1 or Example 2 to obtain a quasistatic approximation for the response of this model to either one or more continuous-wave laser lasers or a mode-locked to simulate the measurements that we have made earlier.

Mode-locking a laser increases the peak intensity in the output which may be seen in Eq. (123), where N is the number of oscillating modes [45]. The output of a mode-locked laser consists of a periodic sequence of short pulses with the period T = 1/Δf. Each pulse has a duration Δt ≈ 2π/NΔω which is equal to the periodicity for the sequence divided by the number of modes. The peak intensity for the output of the laser is proportional to $N^2$. The effective voltage is given by the square-root of this expression. When the laser is focused on a tunneling junction the effective potential energy across the junction is given by Eq. (124) where $E_{x0}$ is the peak value of $E_x$ from the laser at the junction. Thus, in simulating the effect of a mode-locked laser we use Eq. (124) as the effective time-dependent potential.

$$I_{max}(t) \approx I_0 \frac{\sin^2\left(N\frac{\Delta\omega t}{2}\right)}{\sin^2\left(\frac{\Delta\omega t}{2}\right)} \qquad (123)$$

$$U(t) \approx \frac{eE_{x0}a}{N} \frac{\sin\left(N\frac{\Delta\omega t}{2}\right)}{\sin\left(\frac{\Delta\omega t}{2}\right)} \qquad (124)$$

## XI. SUMMARY

We have defined algorithms to determine the wavefunction for electrons that propagate in nanoscale circuits and applied these algorithms in two circuits as examples. There are sharply-defined modes, where each has a specific set of values for the circuit parameters. Figures were prepared using $10^5$ points to enable sufficient resolution to see the zeros throughout a relatively wide search but we recommend that a search algorithm be used to locate the transitions from negative to positive slope followed by bisection to locate each zero in the determinant.

Algorithms are also presented for determining the normalized coefficients for the wavefunction at each mode with the two circuits as examples.

Each mode is determined by setting the determinant of the matrix to zero. Then the normalized coefficients in the wavefunction for each mode may be determined by solving matrix equations which are presented. Then the distribution of current through the circuit at each mode may be determined.

Part V of this paper addresses the issue that in a nanoscale circuit it is too simplistic to assume that a potential barrier has only one incident, one transmitted, and one reflected wave. In Section 4 of the Appendix we have shown how to interpret the transmission from left-to-right and from right-to-left within the barrier for Example 1. This method may be used to define the waves that are generally in both directions in each section of a nanoscale circuit.

## XII. CONCLUSIONS

Our simulations suggest that, because the mean-free path is as large as 68.2 nm in some metals [8], the limited effects of scattering makes it appropriate to solve the Schrödinger equation to determine the wavefunction within the metal when the electrical potential and the energy of the electrons are defined. The wavefunction may be used to determine the probability current density, and thus the electrical current density, within the metal.

We acknowledge that others have developed numerical methods to model nanoscale circuits [46],[47],[48],[49],[50],[51]. The work by Pierantoni, Marcarelli, and Rozzi [51] is especially pertinent to our work because they mention the possibility of ballistic transport within nanoscale devices. However, they do not address the significance of the surprisingly large mean-free path for electrons which has been determined for 20 different metals by Gall [8]. Pierantoni, Marcarelli, and Rozzi [51] refer to measurements showing that the current in CNT FETs may be dramatically altered by small changes in the gate voltage and we suggest that this effect may relate to the coherent transport over long distances which we are studying.

We are considering two types of applications for this new approach: (1) understanding and mitigating unwanted effects that are already seen with present nanoscale circuits [45], and (2) developing new devices as attachments to full-size instruments. Present methods for semiconductor metrology such as scanning capacitance microscopy (SCM) and scanning spreading resistance microscopy (SSRM) use probes to make a 15-nm diameter contact with the semiconductor so their resolution is not adequate to meet the present crisis in the semiconductor industry at and below the 7-nm technology node [7]. This could be addressed by introducing an SFCM attachment to be added to existing scanning tunneling microscopes [52].

**ACKNOWLEDGMENTS**

Dr. Hagmann is grateful to Rolf Landauer and Markus Büttiker who encouraged him to publish his initial work on the numerical modeling of laser-assisted quantum tunneling in 1995, and to Professor Marwan Mousa, at Mu'tah University in Jordan, who made it possible to make time-dependent measurements of laser-assisted field emission during his sabbatical visit to Florida International University from 1999-2000. We acknowledge that Logan Gibb assisted in the graphics as a contractor. We are also grateful to Dmitry Yarotski who made it possible to extend these measurements to laser-assisted Scanning Tunneling Microscopy in visits to the Center for Integrated Nanotechnologies at Los Alamos National Laboratory from 2008 to 2017. Our research on laser-assisted quantum tunneling has been supported by the National Science Foundation under Grant 1648811 and the U.S. Department of Energy under Award DE-SC000639. Our collaboration with the University of Utah was funded by NSF STTR grant 0712564 in 2017 to develop miniature laser-assed field emission devices to generate microwave radiation. The fabrication techniques which we developed at that time [53] will be implemented in our future efforts to prepare prototypes related to Example 2 that is described in this paper.

**APPENDIX**

**Section 1. Static solution of the Schrödinger equation with a linear barrier**

Others have previously solved similar problems using Airy functions [29],[32],[54],[55] but now we present the solution in greater detail because it is used as the first step for the analysis in Part IV of this paper. Figure A1-1 shows the potential energy for a model of quantum tunneling in a static axial electric field. We use the symbol "U" for potential energy to distinguish it from the voltages that are used elsewhere in this paper. A DC electric field $-U_0/a$ causes the potential to decrease linearly over the length of the barrier. We require $U_0 > 0$ and, $0 < E < \phi - U_0$ as shown in this figure so an electron will propagate classically at the left and right of the barrier and tunnel within the full length of the barrier.

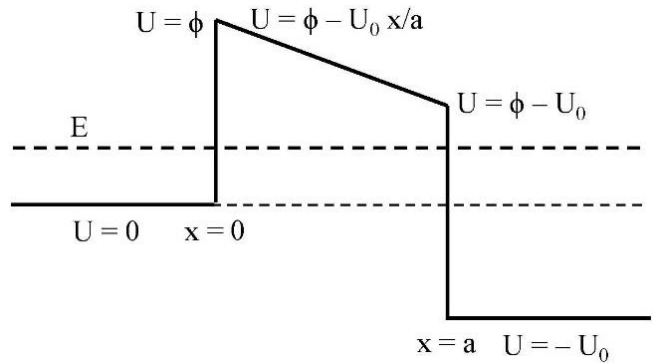

Fig. A1-1. Potential energy for quantum tunneling in a static potential barrier.

With one spatial dimension x, the time-dependent Schrödinger equation is given by Eq. (A1.1). In a static potential this simplifies to Eq. (A1.2) where the wavefunction is given by Eq. (A1.3).

$$\frac{\hbar^2}{2m}\frac{\partial^2 \Psi}{\partial x^2} - U(x,t)\Psi = -i\hbar\frac{\partial \Psi}{\partial t} \qquad (A1.1)$$

$$\frac{\hbar^2}{2m}\frac{d^2\psi}{dx^2}+\left[E-U(x)\right]\psi(x)=0 \tag{A1.2}$$

$$\Psi(x,t)=\psi(x)e^{-i\frac{Et}{\hbar}} \tag{A1.3}$$

**Solve the Schrödinger equation for x < 0.**

In Region 1 to the left of the barrier, the solution of Eq. (A1.3) is given by Eq. (A1.4) for an incident wave with unit amplitude and a reflected wave with coefficient R:

$$\psi_1 = e^{i\sqrt{2mE}\frac{x}{\hbar}} + \mathrm{Re}^{-i\sqrt{2mE}\frac{x}{\hbar}} \tag{A1.4}$$

**Solve the Schrödinger equation for x > a.**

In Region 3, to the right of the barrier, the wavefunction is given by Eq. (A1.5) with only a transmitted wave having the coefficient T:

$$\psi_3 = Te^{i\sqrt{2m(E+U_0)}\frac{x}{\hbar}} \tag{A1.5}$$

**Solve the Schrödinger equation for 0 < x < a.**

In Region 2, within the barrier, the potential energy is given by Eq. (A1.6) and substituting this into Eq. (A1.2) gives Eq. (A1.7).

$$U_2(x) = \phi - U_0\frac{x}{a} \tag{A1.6}$$

$$\frac{\hbar^2}{2m}\frac{d^2\psi_2}{dx^2}+\left[E-\phi+U_0\frac{x}{a}\right]\psi_2=0 \tag{A1.7}$$

A change of variables shown in Eq. (A1.8) is used with Eq. (17) to obtain Eq. (A1.9) where the coefficients A and B have units of meters. Then Eq. (A1.9) is rearranged to obtain Eq. (A1.10).

$$x = A\xi + B \tag{A1.8}$$

$$\frac{\hbar^2}{2mA^2}\frac{d^2\psi_2}{dx^2}+\left[E-\phi+\frac{BU_0}{a}\right]\psi_2+\frac{AU_0}{a}\xi\psi_2=0 \tag{A1.9}$$

$$\frac{d^2\psi_2}{dx^2}+\frac{2mA^2}{\hbar^2}\left[E-\phi+\frac{BU_0}{a}\right]\psi_2+\frac{2mU_0A^3}{\hbar^2 a}\xi\psi_2=0 \tag{A1.10}$$

Next parameter A is chosen by setting the coefficient of the product $\xi\psi_2$ to unity and parameter B is chosen so that the quantity in parentheses in Eq. (A1.10) is zero. Both A and B have units of meters. As noted earlier, $U_0 > 0$ and $E < \phi - U_0$ so that $A > 0$ and $B_1 > a$.

$$A = \left[\frac{\hbar^2 a}{2mU_0}\right]^{\frac{1}{3}} \tag{A1.11}$$

$$B = \left(\frac{\phi - E}{U_0}\right)a \tag{A1.12}$$

Thus, Eq. (A1.10) is simplified to give Eq. (A1.13) which has the solution shown in Eq. (A1.14) where Ai and Bi are Airy functions [29]. The coefficients $C_1$ and $C_2$ have units of inverse meters. Finally, Eq. (A1.8) is used to with Eq. (A1.13) to obtain Eq. (A1.15) where the independent variable is x. Others have used the notation Ai(-x) and Ai(x) in tables for the values of the Airy function to denote the functions for positive or negative arguments where x is a positive number [56]. However, we use the notation that the quantity within the parentheses is simply the argument of the function.

$$\frac{d^2\psi_2}{d\xi^2} + \xi\psi_2 = 0 \tag{A1.13}$$

$$\psi_2(\xi) = C_1 Ai(-\xi) + C_2 Bi(-\xi) \tag{A1.14}$$

$$\psi_2(x) = C_1 Ai\left(\frac{B-x}{A}\right) + C_2 Bi\left(\frac{B-x}{A}\right) \tag{A1.15}$$

Note that when the potential is greater than the energy the parameter B is positive but when the potential is less than the energy B is negative. Thus, the Ai and Bi are monotonically increasing or decreasing functions when the potential exceeds the energy and are oscillatory when the energy exceeds the potential. This is analogous to the exponential and sinusoidal behavior that occur when the potential is a constant. Thus, the Airy functions with complex arguments are not required in these solutions. Furthermore, there is an automatic change from one behavior to the other in a barrier at the point when the potential crosses to be above or below the energy.

**Determine the wavefunction and its derivative at the two boundaries.**

The wavefunction within the barrier has the following values at the ends of the barrier, $x = 0$ and $x = a$, where we note that the argument of the Airy functions is greater than zero.

$$\psi_{2(x=0)} = C_1 Ai\left(\frac{B}{A}\right) + C_2 Bi\left(\frac{B}{A}\right) \tag{A1.16}$$

$$\psi_{2(x=a)} = C_1 Ai\left(\frac{B-a}{A}\right) + C_2 Bi\left(\frac{B-a}{A}\right) \tag{A1.17}$$

The derivative of the wavefunction within the barrier is given by Eq. (A1.18) where Ai' and Bi' are the derivatives of the Ai and Bi functions. Thus, the derivatives at $x = 0$ and $x = a$ are given in Eqs. (A1.19) and (A1.20):

$$\frac{d\psi_2}{dx} = -\frac{C_1}{A} Ai'\left(\frac{B-x}{A}\right) - \frac{C_2}{A} Bi'\left(\frac{B-x}{A}\right) \tag{A1.18}$$

$$\frac{d\psi_2}{dx}_{(x=0)} = -\frac{C_1}{A} Ai'\left(\frac{B}{A}\right) - \frac{C_2}{A} Bi'\left(\frac{B}{A}\right) \tag{A1.19}$$

$$\frac{d\psi_2}{dx}_{(x=a)} = -\frac{C_1}{A} Ai'\left(\frac{B-a}{A}\right) - \frac{C_2}{A} Bi'\left(\frac{B-a}{A}\right) \tag{A1.20}$$

The following expressions for the wavefunctions to the left and right of their barrier and their derivatives at $x = 0$ and $x = a$ are obtained by using Eqs. (A1.4) and (A1.5) for the wavefunctions to the left and right of the barrier.

$$\psi_{1(x=0)} = 1 + R \tag{A1.21}$$

$$\frac{d\psi_1}{dx}_{(x=0)} = \frac{i}{\hbar}\sqrt{2mE} - \frac{i}{\hbar}\sqrt{2mE}\,R \tag{A1.22}$$

$$\psi_{3(x=a)} = Te^{i\sqrt{2m(E+U_0)}\frac{a}{\hbar}} \tag{A1.23}$$

$$\frac{d\psi_3}{dx}_{(x=a)} = \frac{i}{\hbar}\sqrt{2m(E+U_0)}\,Te^{i\sqrt{2m(E+U_0)}\left(\frac{a}{\hbar}\right)} \tag{A1.24}$$

**Apply the boundary conditions to determine the coefficients.**

1. Using Eqs. (A1.16) and (A1.21) for continuity of the wavefunction at x = 0:

$$C_1 Ai\left(\frac{B}{A}\right) + C_2 Bi\left(\frac{B}{A}\right) - R = 1 \qquad (A1.25)$$

2. Using Eqs. (A1.17) and (A1.23) for continuity of the wavefunction at x = a:

$$C_1 Ai\left(\frac{B-a}{A}\right) + C_2 Bi\left(\frac{B-a}{A}\right) - Te^{i\sqrt{2m(E+U_0)}\frac{a}{\hbar}} = 0 \qquad (A1.26)$$

3. Using Eqs. (A1.19) and (A1.22) for continuity of the spatial derivative at x = 0:

$$\frac{C_1}{A} Ai'\left(\frac{B}{A}\right) + \frac{C_2}{A} Bi'\left(\frac{B}{A}\right) - \frac{i}{\hbar}\sqrt{2mE}\, R = -\frac{i}{\hbar}\sqrt{2mE} \qquad (A1.27)$$

4. Using Eqs. (A1.20) and (A1.24) for continuity of the spatial derivative at x = a:

$$\frac{C_1}{A} Ai'\left(\frac{B-a}{A}\right) + \frac{C_2}{A} Bi'\left(\frac{B-a}{A}\right) + \frac{i}{\hbar}\sqrt{2m(E+U_0)}\, Te^{i\sqrt{2m(E+U_0)}\left(\frac{a}{\hbar}\right)} = 0 \qquad (A1.28)$$

To simplify the notation, we define the parameters $P_1$, $P_2$, and $P_3$ in Eqs. (A1.29), (A1.30), and (A1.31).

$$P_1 \equiv \frac{i}{\hbar}\sqrt{2mE} \qquad (A1.29)$$

$$P_2 \equiv \frac{i}{\hbar}\sqrt{2m(E+U_0)} \qquad (A1.30)$$

$$P_3 = e^{i\sqrt{2m(E+U_0)}\left(\frac{a}{\hbar}\right)} \qquad (A1.31)$$

Equations (A1.25), (A1.26), (A1.27), and (A1.28), constitute a system of 4 simultaneous equations in the 4 coefficients $C_1$, $C_2$, R, and T, which has been simplified to give Eqs. (A1.32), (A1.33), (A1.34), and (A1.35).

$$C_1 Ai\left(\frac{B}{A}\right) + C_2 Bi\left(\frac{B}{A}\right) - R = 1 \qquad (A1.32)$$

$$C_1 Ai\left(\frac{B-a}{A}\right) + C_2 Bi\left(\frac{B-a}{A}\right) - P_3 T = 0 \qquad (A1.33)$$

$$C_1 Ai'\left(\frac{B}{A}\right) + C_2 Bi'\left(\frac{B}{A}\right) - AP_1 R = -AP_1 \qquad (A1.34)$$

$$C_1 Ai'\left(\frac{B-a}{A}\right) + C_2 Bi'\left(\frac{B-a}{A}\right) - AP_2 P_3 T = 0 \qquad (A1.35)$$

Solving this system of four equations with four unknown coefficients gives the following expressions for the coefficients $C_1$, $C_2$, R, and T, with D which is the denominator in each of these four equations.

$$C_1 = \frac{2AP_1}{D}\left[Bi'\left(\frac{B-a}{A}\right) - AP_2 Bi\left(\frac{B-a}{A}\right)\right] \qquad (A1.36)$$

$$C_2 = \frac{2AP_1}{D}\left[AP_2 Ai\left(\frac{B-a}{A}\right) - Ai'\left(\frac{B-a}{A}\right)\right] \qquad (A1.37)$$

$$R = \frac{2AP_1}{D}\left[Bi'\left(\frac{B-a}{A}\right) - AP_2 Bi\left(\frac{B-a}{A}\right)\right] Ai\left(\frac{B}{A}\right)$$

$$+ \frac{2AP_1}{D}\left[AP_2 Ai\left(\frac{B-a}{A}\right) - Ai'\left(\frac{B-a}{A}\right)\right] Bi\left(\frac{B}{A}\right) - 1 \quad (A1.38)$$

$$T = \frac{2AP_1}{P_3 D}\left[Bi'\left(\frac{B-a}{A}\right) - AP_2 Bi\left(\frac{B-a}{A}\right)\right] Ai\left(\frac{B-a}{A}\right)$$

$$+ \frac{2AP_1}{P_3 D}\left[AP_2 Ai\left(\frac{B-a}{A}\right) - Ai'\left(\frac{B-a}{A}\right)\right] Bi\left(\frac{B-a}{A}\right) \quad (A1.39)$$

$$D = \left[Bi'\left(\frac{B}{A}\right) - AP_1 Bi\left(\frac{B}{A}\right)\right]\left[Ai'\left(\frac{B-a}{A}\right) - AP_2 Ai\left(\frac{B-a}{A}\right)\right]$$

$$- \left[Ai'\left(\frac{B}{A}\right) - AP_1 Ai\left(\frac{B}{A}\right)\right]\left[Bi'\left(\frac{B-a}{A}\right) - AP_2 Bi\left(\frac{B-a}{A}\right)\right] \quad (A1.40)$$

The probability of tunneling is given by Eq. (A1.41), which is simplified to give Eq. (A1.42).

$$TT^* = \left\{\begin{array}{l}\frac{2AP_1}{P_3 D}\left[Bi'\left(\frac{B-a}{A}\right) - AP_2 Bi\left(\frac{B-a}{A}\right)\right] Ai\left(\frac{B-a}{A}\right) \\ + \frac{2AP_1}{P_3 D}\left[AP_2 Ai\left(\frac{B-a}{A}\right) - Ai'\left(\frac{B-a}{A}\right)\right] Bi\left(\frac{B-a}{A}\right)\end{array}\right\}$$

$$\left\{\begin{array}{l}\frac{2AP_1^*}{P_3 D^*}\left[Bi'\left(\frac{B-a}{A}\right) - AP_2^* Bi\left(\frac{B-a}{A}\right)\right] Ai\left(\frac{B-a}{A}\right) \\ + \frac{2AP_1^*}{P_3 D^*}\left[AP_2^* Ai\left(\frac{B-a}{A}\right) - Ai'\left(\frac{B-a}{A}\right)\right] Bi\left(\frac{B-a}{A}\right)\end{array}\right\} \quad (A1.41)$$

$$TT^* = \frac{4A^2}{P_3^2 DD^*}\left[Ai\left(\frac{B-a}{A}\right) Bi'\left(\frac{B-a}{A}\right) - Ai'\left(\frac{B-a}{A}\right) Bi\left(\frac{B-a}{A}\right)\right]^2 \quad (A1.42)$$

**Section 2. Identity so simplify the summations of Bessel functions in Part IV of the paper**

This identity shown between Eq. (A2.1) and Eq. (A2.8) is used as Eq. (12) in the body of the paper. Consider the summation in Eq. (A2.1) which we group to form Eq. (A2.2) and then use Euler's rule to obtain Eq. (A2.3). Equations (A2.4) and (A2.5) are two identities from [20], which were verified numerically and used to obtain Eq. (A2.6). Then Eq. (A2.6) was simplified to obtain Eq. (A2.7). Finally, Euler's rule is used to obtain Eq. (A2.8), where $\beta \equiv \omega t$, as an alternative expression for the summation S.

$$S \equiv \sum_{-\infty}^{\infty} J_n(\alpha) e^{-in\beta} \quad (A2.1)$$

$$S = J_0(\alpha) - J_1(\alpha)\left[e^{i\beta} - e^{-i\beta}\right] + J_2(\alpha)\left[e^{2i\beta} + e^{-2i\beta}\right]$$
$$- J_3(\alpha)\left[e^{3i\beta} - e^{-3i\beta}\right] + J_4(\alpha)\left[e^{4i\beta} + e^{-4i\beta}\right]$$
$$- J_5(\alpha)\left[e^{5i\beta} - e^{-5i\beta}\right] + J_6(\alpha)\left[e^{6i\beta} + e^{-6i\beta}\right] - \cdots \quad (A2.2)$$

$$S = J_0(\alpha) - 2iJ_1(\alpha)\sin(\beta) + 2J_2(\alpha)\cos(2\beta)$$
$$- 2iJ_3(\alpha)\sin(3\beta) + 2J_4(\alpha)\cos(4\beta)$$
$$- 2iJ_5(\alpha)\sin(5\beta) + 2J_6(\alpha)\cos(6\beta) - \cdots \quad (A2.3)$$

$$\sum_{k=0}^{\infty} J_{2k+1}(\alpha)\sin\left[(2k+1)\beta\right] \equiv \frac{1}{2}\sin\left[\alpha\sin(\beta)\right] \quad (A2.4)$$

$$\sum_{k=1}^{\infty} J_{2k}(\alpha)\cos(2k\beta) \equiv \frac{1}{2}\cos\left[\alpha\sin(\beta)\right] - \frac{1}{2}J_0(\alpha) \quad (A2.5)$$

$$S = J_0(\alpha) + 2\left\{\frac{1}{2}\cos\left[\alpha\sin(\beta)\right] - \frac{1}{2}J_0(\alpha)\right\} - 2i\left\{\frac{1}{2}\sin\left[\alpha\sin(\beta)\right]\right\} \quad (A2.6)$$

$$S = \cos\left[\alpha\sin(\beta)\right] - i\sin\left[\alpha\sin(\beta)\right] \quad (A2.7)$$

$$S = e^{-i\alpha\sin(\beta)} \quad (A2.8)$$

**Section 3. Current density in the solution of the Schrödinger equation for Example 1.**

The electrical current density is given by Eq. (A3.1) for the static solution.

$$J_X(x) = \frac{-ie\hbar}{2m}\left(\psi\frac{d\psi^*}{dx} - \psi^*\frac{d\psi}{dx}\right) \quad (A3.1)$$

Substituting Eqs. (38) and (40) from Part VI of this paper for the wavefunction and its spatial derivative in Region 1 into Eq. (A3.1) we obtain Eq. (A3.2) for the current density in that region. Simplifying Eq. (A3.2) gives Eq. (A3.3), and Eq. (A3.4), where and the current density is given by Eq. (3.5) as the real part of Eq. (3.4).

$$J_{X1}(x) = \frac{-ie\hbar}{2m}\begin{bmatrix}+ik_1\left(C_1e^{-ik_1x} + C_2e^{ik_1x}\right)\left(C_1^*e^{ik_1x} - C_2^*e^{-ik_1x}\right) \\ +ik_1\left(C_1^*e^{ik_1x} + C_2^*e^{-ik_1x}\right)\left(C_1e^{-ik_1x} - C_2e^{ik_1x}\right)\end{bmatrix} \quad (A3.2)$$

$$J_{X1}(x) = \frac{e\hbar k_1}{2m}\begin{bmatrix}\left(C_1e^{-ik_1x} + C_2e^{ik_1x}\right)\left(C_1^*e^{ik_1x} - C_2^*e^{-ik_1x}\right) \\ +\left(C_1^*e^{ik_1x} + C_2^*e^{-ik_1x}\right)\left(C_1e^{-ik_1x} - C_2e^{ik_1x}\right)\end{bmatrix} \quad (A3.3)$$

$$J_{X1}(x) = \frac{e\hbar k_1}{2m}\begin{bmatrix}2C_1C_1^* - 2C_2C_2^* \\ +C_1^*C_2e^{2ik_1x} + C_1C_2^*e^{-2ik_1x} \\ -C_1C_2^*e^{-2ik_1x} - C_1^*C_2e^{2ik_1x}\end{bmatrix} \quad (A3.4)$$

$$J_{X1}(x) = \frac{e\hbar k_1}{m}\left(C_1C_1^* - C_2C_2^*\right) \quad (A3.5)$$

Substituting Eqs. (50) and (51) from Part VI of this paper for the wavefunction and its derivative in Region 2 into Eq. (A3.1) we obtain Eq. (A3.6) for the current density. Simplifying Eq. (A3.6) gives Eq. (3.7) for the current density in Region 2.

$$J_{X2}(x) = \frac{-ie\hbar}{2m} \begin{bmatrix} -\left[C_3 Ai\left(\frac{B_2-x}{A_2}\right) + C_4 Bi\left(\frac{B_2-x}{A_2}\right)\right]\left[\frac{C_3^*}{A_2} Ai'\left(\frac{B_2-x}{A_2}\right)\right] \\ -\left[C_3 Ai\left(\frac{B_2-x}{A_2}\right) + C_4 Bi\left(\frac{B_2-x}{A_2}\right)\right]\left[\frac{C_4^*}{A_2} Bi'\left(\frac{B_2-x}{A_2}\right)\right] \\ +\left[C_3^* Ai\left(\frac{B_2-x}{A_2}\right) + C_4^* Bi\left(\frac{B_2-x}{A_2}\right)\right]\left[\frac{C_3}{A_2} Ai'\left(\frac{B_2-x}{A_2}\right)\right] \\ +\left[C_3^* Ai\left(\frac{B_2-x}{A_2}\right) + C_4^* Bi\left(\frac{B_2-x}{A_2}\right)\right]\left[\frac{C_4}{A_2} Bi'\left(\frac{B_2-x}{A_2}\right)\right] \end{bmatrix} \quad (A3.6)$$

$$J_{X2}(x) = \frac{-ie\hbar}{2mA_2}\left(C_3 C_4^* - C_3^* C_4\right)\begin{bmatrix} Bi\left(\frac{B_2-x}{A_2}\right)Ai'\left(\frac{B_2-x}{A_2}\right) \\ +Ai\left(\frac{B_2-x}{A_2}\right)Bi'\left(\frac{B_2-x}{A_2}\right) \end{bmatrix} \quad (A3.7)$$

Substituting Eqs. (36) and (37) from Part VI of this paper for the wavefunction and its derivative in Region 3 into Eq. (A3.1) we obtain Eq. (A3.8) for the current density, and simplifying Eq. (A3.8) gives Eq. (A3.10) for the current density in Region 3.

$$J_{X3}(x) = \frac{-ie\hbar}{2m} \begin{bmatrix} -\left[C_5 Ai\left(\frac{B_3-x}{A_3}\right) + C_6 Bi\left(\frac{B_3-x}{A_3}\right)\right]\left[\frac{C_5^*}{A_3} Ai'\left(\frac{B_3-x}{A_3}\right)\right] \\ -\left[C_5 Ai\left(\frac{B_3-x}{A_3}\right) + C_6 Bi\left(\frac{B_3-x}{A_3}\right)\right]\left[\frac{C_6^*}{A_3} Bi'\left(\frac{B_3-x}{A_3}\right)\right] \\ +\left[C_5^* Ai\left(\frac{B_3-x}{A_3}\right) + C_6^* Bi\left(\frac{B_3-x}{A_3}\right)\right]\left[\frac{C_5}{A_3} Ai'\left(\frac{B_3-x}{A_3}\right)\right] \\ +\left[C_5^* Ai\left(\frac{B_3-x}{A_3}\right) + C_6^* Bi\left(\frac{B_3-x}{A_3}\right)\right]\left[\frac{C_6}{A_3} Bi'\left(\frac{B_3-x}{A_3}\right)\right] \end{bmatrix} \quad (A3.8)$$

$$J_{X3}(x) = \frac{-ie\hbar}{2mA_3}\left(C_5 C_6^* - C_5^* C_6\right)\begin{bmatrix} Bi\left(\frac{B_3-x}{A_3}\right)Ai'\left(\frac{B_3-x}{A_3}\right) \\ -Ai\left(\frac{B_3-x}{A_3}\right)Bi'\left(\frac{B_3-x}{A_3}\right) \end{bmatrix} \quad (A3.10)$$

**Section 4. Interpretation in terms of forward and reflected waves in Example 1.**

The wavefunction within the barrier is given by Eq. (50) for Example 1 in Part VI of this paper, which we copy as Eq. (A4.1) for convenience of the reader:

$$\psi_2(x) = C_3 Ai\left(\frac{B_2-x}{A_2}\right) + C_4 Bi\left(\frac{B_2-x}{A_2}\right) \quad (A4.1)$$

Thus, at x = 0, and x = a,

$$\psi_2(0) = C_3 Ai\left(\frac{B_2}{A_2}\right) + C_4 Bi\left(\frac{B_2}{A_2}\right) \quad (A4.2)$$

$$\psi_2(a) = C_3 Ai\left(\frac{B_2 - a}{A_2}\right) + C_4 Bi\left(\frac{B_2 - a}{A_2}\right) \quad (A4.3)$$

When the argument of the two Airy functions is positive, and increasing, Ai has quasi-exponential decay and Bi has quasi-exponential growth. However, if E > ϕ -$U_0$, the argument is zero at x = $B_2$ where the energy E is equal to the potential U, and both Ai and Bi have quasi-sinusoidal behavior for x > $B_2$ because the argument is negative. Thus, in Eq. (A4.1), the term with Bi as the Airy function corresponds to the wave entering the barrier at x = 0 and exiting at x = a, and the term with Ai as the Airy function corresponds to the wave entering the barrier at x = a and exiting at x = 0. Thus, the transmission of the waves traveling from left to right (LR) and from right to left (RL) are given in Eq. (A4.4) and Eq. (A4.5):

$$T_{LR} = \left[\frac{Bi\left(\frac{B_2 - a}{A_2}\right)}{Bi\left(\frac{B_2}{A_2}\right)}\right]^2 \quad (A4.4)$$

$$T_{RL} = \left[\frac{Ai\left(\frac{B_2}{A_2}\right)}{Ai\left(\frac{B_2 - a}{A_2}\right)}\right]^2 \quad (A4.5)$$

We use coefficients $C_{3N} \equiv C_{3R} + iC_{3I}$, and $C_{4N} \equiv C_{4R} + iC_{4I}$, that are normalized such that $C_{1R}$ = 1 and $C_{1I}$ = 0. Thus, the relative magnitude of the two parts of the normalized probability density $\psi\psi^*$ at the entrance and exit when entering and leaving at each side of the barrier are given by the following expressions:

For the wave traveling from left to right in the barrier, Region 2:

$$\text{Entering at x = 0, } \psi\psi^* = \left(C_{4R}^2 + C_{4I}^2\right)\left[Bi\left(\frac{B_2}{A_2}\right)\right]^2 \quad (A4.6)$$

$$\text{Leaving at x = a, } \psi\psi^* = \left(C_{4R}^2 + C_{4I}^2\right)\left[Bi\left(\frac{B_2 - a}{A_2}\right)\right]^2 \quad (A4.7)$$

For the wave traveling from right to left in the barrier, Region 2:

$$\text{Entering at x = a, } \psi\psi^* = \left(C_{3R}^2 + C_{3I}^2\right)\left[Ai\left(\frac{B_2 - a}{A_2}\right)\right]^2 \quad (A4.8)$$

$$\text{Leaving at x = 0, } \psi\psi^* = \left(C_{3R}^2 + C_{3I}^2\right)\left[Ai\left(\frac{B_2}{A_2}\right)\right]^2 \quad (A4.9)$$

**Section 5. Determine the normalized coefficients for Example 1.**

We consider the system of 6 simultaneous equations, Eqs. (96) to (101) in the body of this paper with the normalized coefficients for Example 1. The objective is to determine expressions for the normalized coefficients $C_{2N}$, $C_{3N}$, $C_{4N}$, $C_{5N}$, and $C_{6N}$. Generally, this would be done numerically by solving the matrix equation. However, now we rearrange Eq. (96) to solve for $C_{2N}$ gives Eq. (A5.1), which is then inserted to remove $C_{2N}$ from the remaining 5 equations in this group. Next, rearranging Eq. (98) to give Eq. (A5.2) which is then inserted to remove $C_{3N}$ from the remaining 4 equations in this group. Then Eq. (99) is rearranged to give Eq. (A5.3) which is inserted to remove $C_{4N}$ from the remaining equations.

$$C_{2N} = -\left(M_{13}C_{3N} + M_{14}C_{4N} + 1\right) \tag{A5.1}$$

$$C_{3N} = -\left(M_{34}C_{4N} + M_{35}C_{5N} + M_{36}C_{6N}\right)\frac{1}{M_{33}} \tag{A5.2}$$

$$C_{4N} = -\frac{\left(M_{33}M_{45} - M_{35}M_{43}\right)}{\left(M_{33}M_{44} - M_{34}M_{43}\right)}C_{5N} - \frac{\left(M_{33}M_{46} - M_{36}M_{43}\right)}{\left(M_{33}M_{44} - M_{34}M_{43}\right)}C_{6N} \tag{A5.3}$$

These operations result in Eqs. (A5.4), (A5.5), and (A5.6) which contain only the remaining normalized coefficients $C_{5N}$ and $C_{6N}$.

$$\begin{bmatrix} M_{13}M_{22}M_{34}M_{33}M_{45} - M_{13}M_{22}M_{35}M_{33}M_{44} \\ -M_{14}M_{22}M_{33}M_{33}M_{45} + M_{14}M_{22}M_{33}M_{35}M_{43} \\ -M_{23}M_{34}M_{33}M_{45} + M_{23}M_{34}M_{35}M_{43} + M_{23}M_{35}M_{33}M_{44} \\ -M_{23}M_{35}M_{34}M_{43} + M_{24}M_{33}M_{33}M_{45} - M_{24}M_{33}M_{35}M_{43} \end{bmatrix} C_{5N}$$

$$+ \begin{bmatrix} M_{13}M_{22}M_{34}M_{33}M_{46} - M_{13}M_{22}M_{34}M_{36}M_{43} \\ -M_{13}M_{22}M_{36}M_{33}M_{44} + M_{13}M_{22}M_{36}M_{34}M_{43} \\ -M_{14}M_{22}M_{33}M_{33}M_{46} + M_{14}M_{22}M_{33}M_{36}M_{43} \\ -M_{23}M_{34}M_{33}M_{46} + M_{23}M_{34}M_{36}M_{43} + M_{23}M_{36}M_{33}M_{44} \\ -M_{23}M_{36}M_{34}M_{43} + M_{24}M_{33}M_{33}M_{46} - M_{24}M_{33}M_{36}M_{43} \end{bmatrix} C_{6N}$$

$$= -\left(M_{22}M_{33} - M_{21}M_{33}\right)\left(M_{33}M_{44} - M_{34}M_{43}\right) \tag{A5.4}$$

$$\begin{bmatrix} -M_{13}M_{34}M_{33}M_{45}M_{52} + M_{13}M_{34}M_{35}M_{43}M_{52} + M_{13}M_{35}M_{33}M_{44}M_{52} \\ -M_{13}M_{35}M_{34}M_{43}M_{52}M_{14} + M_{33}M_{33}M_{45}M_{52} - M_{14}M_{33}M_{35}M_{43}M_{52} \\ +M_{33}M_{33}M_{44}M_{55} - M_{33}M_{34}M_{43}M_{55} \end{bmatrix} C_{5N}$$

$$+ \begin{bmatrix} -M_{13}M_{34}M_{33}M_{45}M_{52} + M_{13}M_{34}M_{35}M_{43}M_{52} + M_{13}M_{36}M_{33}M_{44}M_{52} \\ -M_{13}M_{36}M_{34}M_{43}M_{52} + M_{14}M_{33}M_{33}M_{45}M_{52} - M_{14}M_{33}M_{35}M_{43}M_{52} \\ +M_{33}M_{33}M_{44}M_{56} - M_{33}M_{34}M_{43}M_{56} \end{bmatrix} C_{6N}$$

$$= -\left(M_{33}M_{51} - M_{33}M_{52}\right)\left(M_{33}M_{44} - M_{34}M_{43}\right) \tag{A5.5}$$

$$\begin{bmatrix} M_{13}M_{34}M_{33}M_{45}M_{62} - M_{13}M_{34}M_{35}M_{43}M_{62} - M_{13}M_{35}M_{33}M_{44}M_{62} \\ +M_{13}M_{35}M_{34}M_{43}M_{62} - M_{14}M_{33}M_{33}M_{45}M_{62} + M_{14}M_{33}M_{35}M_{43}M_{62} \\ -M_{33}M_{33}M_{44}M_{65} + M_{33}M_{34}M_{43}M_{65} \end{bmatrix} C_{5N}$$

$$+ \begin{bmatrix} M_{13}M_{34}M_{33}M_{46}M_{62} - M_{13}M_{34}M_{36}M_{43}M_{62} - M_{13}M_{36}M_{33}M_{44}M_{62} \\ +M_{13}M_{36}M_{34}M_{43}M_{62} - M_{14}M_{33}M_{33}M_{46}M_{62} + M_{14}M_{33}M_{36}M_{43}M_{62} \\ -M_{33}M_{33}M_{44}M_{66} + M_{33}M_{34}M_{43}M_{66} \end{bmatrix} C_{6N}$$

$$= -(M_{33}M_{62} - M_{33}M_{61})(M_{33}M_{44} - M_{34}M_{43}) \quad (A5.6)$$

We write Eqs. (A5.4), (A5.5), and (A5.6) where A, B, D, E, G, and H are matrices and C, F, and I represent the expressions which are on the RHS of Eqs. (A5.4), (A5.5), and (A5.6), respectively.

$$AC_{5N} + BC_{6N} = C \quad (A5.7)$$
$$DC_{5N} + EC_{6N} = F \quad (A5.8)$$
$$GC_{5N} + HC_{6N} = I \quad (A5.9)$$

The following three expressions are obtained for the remaining normalized coefficients, $C_{5N}$ and $C_{6N}$, by combining Eqs. (A5.7 and A5.8), (A5.9) and (A5.10), and (A5.11) with (A5.12). Each of the three expressions for $C_{5N}$ are in agreement as are each of the three expressions for $C_{6N}$ because of the degeneracy of the system of equations.

Combining Eqs. (A5.7) and (A5.8):

$$C_{5N} = \frac{(CE - BF)}{(AE - BD)} \quad (A5.10)$$

$$C_{6N} = \frac{(AF - CD)}{(AE - BD)} \quad (A5.11)$$

Combining Eqs. (A5.7) and (A5.9):

$$C_{5N} = \frac{(CH - BI)}{(AH - BG)} \quad (A5.12)$$

$$C_{6N} = \frac{(AI - CG)}{(AH - BG)} \quad (A5.13)$$

Combining Eqs. (A5.8) and (A5.9):

$$C_{5N} = \frac{EI - FH}{(EG - DH)} \quad (A5.14)$$

$$C_{6N} = \frac{(FG - DI)}{(EG - DH)} \quad (A5.15)$$

**Section 6. Solution in Region 2 for Example 2 in Part X of the body of this paper.**

Equation (A6.1) is obtained by using Eq. (109) from Part X in the body of this paper as the potential in the Schrödinger equation.

$$\frac{\hbar^2}{2m}\frac{d^2\psi_2}{dx^2}+\left[E+eE_x(t)\frac{a}{2}-eE_xx\right]\psi_2=0 \quad (A6.1)$$

A change of variables that is shown as Eq. (A6.2) is used in Eq. (A6.1) to obtain Eq. (A6.3) which is rearranged to obtain Eq. (129).

$$x = A_2\xi + B_2 \quad (A6.2)$$

$$\frac{d^2\psi_2}{d\xi^2}+\frac{2mA_2^2}{\hbar^2}\left[E+eE_x\frac{a}{2}-eE_xB_2\right]\psi_2-\frac{2m}{\hbar^2}eE_xA_2^3\xi\psi_2=0 \quad (A6.3)$$

Parameter $B_2$ is set as shown in Eq. (A6.4) so that the second term in Eq. (A6.3) is zero and the parameter $A_2$ is chosen as shown in Eq. (A6.5) so that the coefficient of $\xi\psi_2$ in the third term of Eq. (A6.3) is unity. This gives the standard form for the Airy equation in Eq. (A6.6) [29] for which the solution is given in Eq. (A6.7). Finally, inserting the two coefficients that are given in Eqs. (A6.4) and (A6.5) results in the solution for the static case of the wavefunction $\psi_2$ as Eq. (A6.8).

$$B_2 = \frac{a}{2}+\frac{E}{eE_x} \quad (A6.4)$$

$$A_2 = -\left(\frac{\hbar^2}{2meE_x}\right)^{\frac{1}{3}} \quad (A6.5)$$

$$\frac{d^2\psi_2}{d\xi^2}+\xi\psi_2=0 \quad (A6.6)$$

$$\psi_2 = F_2A_i(-\xi)+G_2B_i(-\xi) \quad (A6.7)$$

$$\psi_2 = F_2A_i\left(\frac{B_2-x}{A_2}\right)+G_2B_i\left(\frac{B_2-x}{A_2}\right) \quad (A6.8)$$

The transition from positive to negative arguments in the two Airy functions in Eq. (A6.8) occurs at $x = B_2$, which equals $X_C$ in Eq. (111) defined in the body of this paper. This is the point of the transition between quasi-exponential and quasi-sinusoidal behavior for each of the two Airy functions. Thus, in Region 2 the transition between quantum tunneling and classical transport within the barrier is built into this part of the solution without requiring the use of two separate parts as is done in Region 1 and Region 3.

**Section 7. Apply the boundary conditions for Example 2 in Part X of this paper.**

Depending on the value of the energy and the potentials, there are four separate cases for the wavefunctions in the three regions when using Eqs. (113), (115), (117), (119), (122) from Part X of this paper for the wavefunctions:

Case 1 requires that $E > U_1$ and $U_3$, using the parameters $k_{1A}$, $\gamma_2$, and $k_{3A}$ in Eqs. (113), (117) and (122).

Case 2 requires that $U_1 < E < U_3$, using the parameters $k_{1A}$, $\gamma_2$, and $\gamma_{3B}$ in Eqs. (113), (119) and (122).

Case 3 requires that $U_3 < E < U_1$, using the parameters $\gamma_{1B}$, $\gamma_2$, and $k_{3A}$ in Eqs. (115), (117) and (122).

Case 4 requires that $E < U_1$ and $U_3$, using the parameters $\gamma_{1B}$, $\gamma_2$, and $\gamma_{3B}$ in Eqs. (115), (119) and (122).

**For Case 1**

1. Requiring the wavefunction has a null at the left end of the model where x = -d:
$$A_{1A}e^{ik_{1A}d} + B_{1A}e^{-ik_{1A}d} = 0 \quad (A7.1)$$

2. Requiring the wavefunction has a null at right end of the model where x = a +d:
$$A_{3A}e^{-ik_{3A}(a+d)} + B_{3A}e^{ik_{3A}(a+d)} = 0 \quad (A7.2)$$

3. Requiring the wavefunction to be continuous at x = 0:
$$F_2 A_i\left(\frac{B_2}{A_2}\right) + G_2 B_i\left(\frac{B_2}{A_2}\right) - A_{1A} - B_{1A} = 0 \quad (A7.3)$$

4. Requiring the spatial derivative of the wavefunction to be continuous at x = 0:
$$\frac{F_2}{A_2} A_i{}'\left(\frac{B_2}{A_2}\right) + \frac{G_2}{A_2} B_i{}'\left(\frac{B_2}{A_2}\right) - ik_{1A}A_{1A} + ik_{1A}B_{1A} = 0 \quad (A7.4)$$

5. Requiring the wavefunction to be continuous at x = a:
$$F_2 A_i\left(\frac{B_2 - a}{A_2}\right) + G_2 B_i\left(\frac{B_2 - a}{A_2}\right) - A_{3A}e^{-ik_{3A}a} - B_{3A}e^{ik_{3A}a} = 0 \quad (A7.5)$$

6. Requiring the spatial derivative of the wavefunction to be continuous at x = a:
$$\frac{F_2}{A_2} A_i{}'\left(\frac{B_2 - a}{A_2}\right) \frac{G_2}{A_2} B_i{}'\left(\frac{B_2 - a}{A_2}\right) - ik_{3A}A_{3A}e^{-ik_{3A}a} + ik_{3A}B_{3A}e^{ik_{3A}a} = 0 \quad (A7.6)$$

**For Case 2**

1. Require the wavefunction has a null at the left end of the model where x = -d.
$$A_{1A}e^{ik_{1A}d} + B_{1A}e^{-ik_{1A}d} = 0 \quad (A7.7)$$

2. Require the wavefunction has a null at right end of the model where x = a +d.
$$C_{3B}e^{-\gamma_{3B}(a+d)} + D_{3B}e^{\gamma_{3B}(a+d)} = 0 \quad (A7.8)$$

3. Require the wavefunction to be continuous at x = 0.
$$F_2 A_i\left(\frac{B_2}{A_2}\right) + G_2 B_i\left(\frac{B_2}{A_2}\right) - A_{1A} - B_{1A} = 0 \quad (A7.9)$$

4. Require the spatial derivative of the wavefunction to be continuous at x = 0.
$$\frac{F_2}{A_2} A_i{}'\left(\frac{B_2}{A_2}\right) + \frac{G_2}{A_2} B_i{}'\left(\frac{B_2}{A_2}\right) - ik_{1A}A_{1A} + ik_{1A}B_{1A} = 0 \quad (A7.10)$$

5. Require the wavefunction to be continuous at x = a.
$$F_2 A_i\left(\frac{B_2 - a}{A_2}\right) + G_2 B_i\left(\frac{B_2 - a}{A_2}\right) - C_{3B}e^{-\gamma_{3B}a} - D_{3B}e^{\gamma_{3B}a} = 0 \quad (A7.11)$$

6. Require the spatial derivative of the wavefunction to be continuous at x = a.
$$\frac{F_2}{A_2} A_i\left(\frac{B_2 - a}{A_2}\right) + \frac{G_2}{A_2} B_i\left(\frac{B_2 - a}{A_2}\right) - \gamma_{3B}C_{3B}e^{-\gamma_{3B}a} + \gamma_{3B}D_{3B}e^{\gamma_{3B}a} = 0 \quad (A7.12)$$

**For Case 3**

1. Require the wavefunction has a null at the left end of the model where x = -d.

$$C_{1B}e^{\gamma_{1B}d} + D_{1B}e^{-\gamma_{1B}d} = 0 \tag{A7.13}$$

2. Require the wavefunction has a null at right end of the model where x = a +d.
$$A_{3A}e^{-ik_{3A}(a+d)} + B_{3A}e^{ik_{3A}(a+d)} = 0 \tag{A7.14}$$

3. Require the wavefunction to be continuous at x = 0.
$$F_2 A_i\left(\frac{B_2}{A_2}\right) + G_2 B_i\left(\frac{B_2}{A_2}\right) - C_{1B} - D_{1B} = 0 \tag{A7.15}$$

4. Require the spatial derivative of the wavefunction to be continuous at x = 0.
$$\frac{1}{A_2} A_i{'}\left(\frac{B_2}{A_2}\right) F_2 + \frac{1}{A_2} B_i{'}\left(\frac{B_2}{A_2}\right) G_2 - \gamma_{1B} C_{1B} + \gamma_{1B} D_{1B} = 0 \tag{A7.16}$$

5. Require the wavefunction to be continuous at x = a.
$$F_2 A_i\left(\frac{B_2 - a}{A_2}\right) + G_2 B_i\left(\frac{B_2 - a}{A_2}\right) - A_{3A}e^{-ik_{3A}a} - B_{3A}e^{ik_{3A}a} = 0 \tag{A7.17}$$

6. Require the spatial derivative of the wavefunction to be continuous at x = a.
$$\frac{1}{A_2} A_i{'}\left(\frac{B_2 - a}{A_2}\right) F_2 + \frac{1}{A_2} B_i{'}\left(\frac{B_2 - a}{A_2}\right) G_2 - ik_{3A} A_{3A} e^{-ik_{3A}a} + ik_{3A} B_{3A} e^{ik_{3A}a} = 0 \tag{A7.18}$$

**For Case 4**

1. Require the wavefunction has a null at the left end of the model where x = -d.
$$C_{1B}e^{\gamma_{1B}d} + D_{1B}e^{-\gamma_{1B}d} = 0 \tag{A7.19}$$

2. Require the wavefunction has a null at right end of the model where x = a +d.
$$C_{3B}e^{-\gamma_{3B}(a+d)} + D_{3B}e^{\gamma_{3B}(a+d)} = 0 \tag{A7.20}$$

3. Require the wavefunction to be continuous at x = 0.
$$F_2 A_i\left(\frac{B_2}{A_2}\right) + G_2 B_i\left(\frac{B_2}{A_2}\right) - C_{1B} - D_{1B} = 0 \tag{A7.21}$$

4. Require the spatial derivative of the wavefunction to be continuous at x = 0.
$$\frac{F_2}{A_2} A_i{'}\left(\frac{B_2}{A_2}\right) + \frac{G_2}{A_2} B_i{'}\left(\frac{B_2}{A_2}\right) - \gamma_{1B} C_{1B} + \gamma_{1B} D_{1B} = 0 \tag{A7.22}$$

5. Require the wavefunction to be continuous at x = a.
$$F_2 A_i\left(\frac{B_2 - a}{A_2}\right) + G_2 B_i\left(\frac{B_2 - a}{A_2}\right) - C_{3B}e^{-\gamma_{3B}a} - D_{3B}e^{\gamma_{3B}a} = 0 \tag{A7.23}$$

6. Require the spatial derivative of the wavefunction to be continuous at x = a.
$$\frac{F_2}{A_2} A_i{'}\left(\frac{B_2 - a}{A_2}\right) + \frac{G_2}{A_2} B_i{'}\left(\frac{B_2 - a}{A_2}\right) - \gamma_{3B} C_{3B} e^{-\gamma_{3B}a} + \gamma_{3B} D_{3B} e^{\gamma_{3B}a} = 0 \tag{A7.24}$$

In each of the four cases the corresponding set of six homogeneous equations in the six coefficients must be satisfied for the solution to be consistent with the boundary conditions.